\documentclass[12pt,preprint]{aastex6}

\usepackage{amsmath}
\usepackage{natbib}
\usepackage{xcolor}
\usepackage[]{graphics}
\usepackage{epsfig}
\usepackage{epstopdf}

\usepackage{mathrsfs}
\usepackage{appendix}

\begin{document}

\title{The eccentricity enhancement effect of intermediate-mass-ratio-inspirals: dark matter and black hole mass}

\author{
   Meirong Tang \altaffilmark{1,2,3,4}
  Jiancheng Wang \altaffilmark{1,2,3,4}
 }

\altaffiltext{1}{Yunnan Observatories, Chinese Academy of Sciences, 396 Yangfangwang, Guandu District, Kunming, 650216, P. R. China; {\tt  mrtang@ynao.ac.cn,  jcwang@ynao.ac.cn
}}
\altaffiltext{2}{Key Laboratory for the Structure and Evolution of Celestial Objects, Chinese Academy of Sciences, 396 Yangfangwang, Guandu District, Kunming, 650216, P. R. China}
\altaffiltext{3}{Center for Astronomical Mega-Science, Chinese Academy of Sciences, 20A Datun Road, Chaoyang District, Beijing, 100012, P. R. China}
\altaffiltext{4}{University of Chinese Academy of Sciences, Beijing, 100049, P. R. China}

\shorttitle{The eccentricity enhancement effect of IMRI}
\shortauthors{Tang et al.}

\begin{abstract}

It was found that the dark matter (DM) in the intermediate-mass-ratio-inspiral (IMRI) system has a significant enhancement effect on the orbital eccentricity of the stellar massive compact object, such as a black hole (BH), which may be tested by space-based gravitational wave (GW) detectors including LISA, Taiji and Tianqin in future observations \citep{2019PhRvD.100d3013Y}. In this paper, we will study the enhancement effect of the eccentricity for an IMRI under different DM density profiles and center BH masses. Our results are as follows: $(1)$ in terms of the general DM spike distribution, the enhancement of the eccentricity is basically consistent with the power-law profile, which indicates that it is reasonable to adopt the power-law profile; $(2)$ in the presence of DM spike, the different masses of the center BH will affect the eccentricity, which provides a new way for us to detect the BH's mass; $(3)$ considering the change of the eccentricity in the presence and absence of DM spike, we find that it is possible to distinguish DM models by measuring the eccentricity at the scale of about $10^{5} GM/c^{2}$. 

\end{abstract}


\section{INTRODUCTION}
\label{intro}

Dark matter (DM) is one of the major difficulties faced by modern astronomy and physics. Although there are many indirect evidences in its observation, such as cosmic microwave background radiation (CMB), rotation curves (RC) of spiral galaxies, mass-to-light ratio of elliptical galaxies, measurement of large-scale structure of the universe and so on, but physicists and astronomers have so far been unable to directly detect DM particles, or even theoretically what constitutes them \citep{2005PhR...405..279B, 2016A&A...594A..13P}. The problem of DM still faces great challenges and opportunities. For example, the cold dark matter (CDM) model that attracts the most attention from researchers has small-scale observation problems, which mainly include: the missing satellites problem (MSP) \citep{1999ApJ...522...82K, 1999ApJ...524L..19M, 2017ARA&A..55..343B}, the cusp-core problem (CCP) \citep{1994ApJ...427L...1F, 1994Natur.370..629M, 2015A&A...578A..13K, 2017ARA&A..55..343B, 2019MNRAS.483..289R}, the rotation curve diversity problem (RCD) \citep{2015MNRAS.452.3650O, 2017PhRvL.119k1102K}, etc. In recent years, using the measurement of gravitational wave (GW) \citep{2006ApJ...640..156G, 2013PhRvL.110v1101E, 2015PhRvD..91d4045E, 2018PhRvD..97f4003Y, 2018PhRvD..98b3536K, 2019PhRvD.100d3013Y} and black hole shadow (BHS) \citep{2012PhRvD..86j3001Y, 2018JCAP...10..046X, 2018PhRvD..98d4053C, 2018JCAP...12..040H}, scientists have been able to study the properties of DM near black hole (BH), which has opened a new way for DM detection.  

Therefore, it is particularly important to understand the effect of a BH on the distribution of DM. We know that there is a innermost stable circular orbit (ISCO) near the intermediate mass black hole (IMBH) \citep{2013PhRvL.110v1101E, 2015PhRvD..91d4045E}. When the DM particles are within the radius of the ISCO, the DM will not be able to form a stable distribution. The existence of BH makes the DM distribution around the BH appear a density cusp, which is the famous DM spike phenomenon \citep{1999PhRvL..83.1719G, 2019ApJ...874...34Y}. Using the adiabatic approximation and Newton's approximation, \cite{1999PhRvL..83.1719G} studied the DM distribution under the Schwarzchild BH and found that the distribution form was $\rho_{\rm{DM}}(r)=\rho_{\rm{sp}}(1-8GM/c^{2}r)^{3}(r_{\rm{sp}}/r)^{\alpha}$. Meanwhile, they also proposed that the DM distribution around the BH can be replaced by the power-law profile, and the results show that the DM density is zero at $8GM/c^{2}$ (i.e., $4R_{s}$, the Schwarzchild radius $R_{s}=2GM/c^{2}$). In the case of adiabatic approximation, \cite{2013PhRvD..88f3522S} obtained different results from \cite{1999PhRvL..83.1719G} using rigorous general relativity (GR). They found that the DM density was zero at $4GM/c^{2}$(i.e., $2R_{s}$) and increased by more than 15 percent at the peak of the spike, which may cause observable effects in gravitational wave events of the intermediate-mass-ratio inspiral (IMRI) system.   

The existence of BH can greatly increase the density of DM. If DM annihilates into gamma-ray photons, it will enhance the possibility of detecting DM signals. On the other hand, the DM near the BH may have significant dynamical effects, including two aspects. One is the effect of DM on the stellar orbital dynamic. It was found that the effect is very small and there is no observable effect. The other is the effect of DM on gravitational wave (GW) signals generated during the merger of the compact binary. \cite{2013PhRvL.110v1101E} found that the DM spike around the center IMBH has a significant impact on the GW signals of the IMRI system, which indicates that the observation of the GW from IMRI may be of great help in the exploration of DM models. Recently, \cite{2019PhRvD.100d3013Y} studied the enhancement of eccentricity and GW signals for IMRI by DM spike, but they did not study the changes of eccentricity due to the differences of the center IMBH masses and DM models. And they also assumed that the DM spike satisfies a power law profile. In fact, the DM spike could meet a more rigorous distribution. In this work, we will focus on these two issues.

The outline of this paper is as follows. In Sec. \ref{two} we present the DM distribution with spike and without spike. In Sec. \ref{dynamical equations} we study the enhancement effect of eccentricity for an IMRI with different center IMBH masses and DM distributions. The summary and conclusions are given in Sec. \ref{conclusion}.

$$
$$
\section{DM density profiles}
\label{two}

(1) DM profile with spike

According to the reference \cite{1999PhRvL..83.1719G}, when Newton approximation and adiabatic approximation are considered, the distribution of DM spike is $ \rho_{\rm{DM}}(r)=\rho_{\rm{sp}}(1-8GM/c^{2}r)^{3}(r/r_{\rm{sp}})^{-\alpha}$. When considering the general relativity (GR) and adiabatic approximation, \cite{2013PhRvD..88f3522S} got the distribution of DM spike as $ \rho_{\rm{DM}}(r)=\rho_{\rm{sp}}(1-4GM/c^{2}r)^{3}(r/r_{\rm{sp}})^{-\alpha}$. Comprehensively considering the influence of relativity, we proposed a more general distribution of DM spike, which is expressed as

\begin{equation}
\rho_{\rm{DM}}(r)=\rho_{\rm{sp}}(1-\dfrac{kGM}{c^{2}r})^{3}(\dfrac{r_{\rm{sp}}}{r})^{\alpha} ,
\label{New_DM_profile}
\end{equation}
where the $r_{\rm{sp}}$ is the radius of the DM spike, and $\rho_{\rm{sp}}$ is the DM density at the radius $r_{\rm{sp}}$. $M$ is the mass of the IMBH, $k$ is a constant and $4\leq k\leq8$ ($k$ is equal to $4$ in the GR case, and $k$ is equal to $8$ in the Newton approximation case ) \citep{1999PhRvL..83.1719G, 2017arXiv170808449N, 2019arXiv190611845H, 2013PhRvD..88f3522S}, in this work, the results are the same whether $k$ is $4$ or $8$, so we take $k=4$. $G$ is the gravitational constant, and $c$ is the speed of light. $\alpha$ is the power law index of the DM spike, and in this paper we assume $1.5\leq \alpha\leq 7/3$ \citep{1997ApJ...490..493N, 2001PhRvD..64d3504U, 1995ApJ...440..554Q}. According to the reference \citep{2015PhRvD..91d4045E, 2019PhRvD.100d3013Y}, we set $r_{\rm{sp}}=0.54 \rm{pc}$ and $\rho_{\rm{sp}}=226M_{\odot}/ \rm{pc}^{3}$.

(2) DM halo without spike

(i) Navarro-Frenk-White (NFW) density profile

Based on the cosmological constant plus cold dark matter ($\Lambda$CDM) model and numerical simulation \citep{1991ApJ...378..496D, 1996ApJ...462..563N, 1997ApJ...490..493N}, an approximate analytical expression of NFW density profile is derived and expressed as

\begin{equation}
\rho_{\rm{NFW}}(r)=\dfrac{\rho_{\rm{0}}}{\dfrac{r}{R_{\rm{s}}}(1+\dfrac{r}{R_{\rm{s}}})^{2}},
\label{rho_NFW}
\end{equation}
where $\rho_{0}$ is the DM density when the DM halo collapses, and $R_{s}$ is the scale radius.

(ii) Thomas-Fermi (TF) density profile

Based on the Bose-Einstein condensation dark matter (BEC-DM) model and TF approximation \citep{2007JCAP...06..025B}, the DM density profile is given by

\begin{equation}
\rho_{\rm{TF}}(r)=\rho_{\rm{0}}\dfrac{\sin(kr)}{kr},
\label{rho_TF}
\end{equation}
where $\rho_{0}$ is the center density of BEC-DM halo, $k=\pi/R$, $R$ is the radius where the DM pressure and density vanish.

(iii) Pseudo-Isothermal (PI) density profile

Based on the modified Newtonian dynamics (MOND) model \citep{1991MNRAS.249..523B}, the DM density profile is given by

\begin{equation}
\rho_{\rm{PI}}(r)=\dfrac{\rho_{\rm{0}}}{1+(\dfrac{r}{R_{\rm{c}}})^{2}},
\label{rho_PI}
\end{equation}
where $\rho_{0}$ is the center density of DM, $R_{c}$ is the scale radius.

In this work, we set $R_{\rm{s}}$, $R$ and $R_{\rm{c}}$ as $0.54 \rm{pc}$, and $\rho_{0}=226M_{\odot}/ \rm{pc}^{3}$ \citep{2015PhRvD..91d4045E, 2019PhRvD.100d3013Y}.

\section{Dynamical equations and eccentricity enhancement effect for IMRI}
\label{dynamical equations}

\subsection{DM profile with spike} 
\label{3-1}

For a binary system which include a small compact object and an IMBH, if the mass of the small compact object is much less than the mass of the IMBH, then this binary system can be reduce to the small compact object moves in the gravitational field of the center IMBH. Next, we use the same method as \citep{2019PhRvD.100d3013Y} to derive the dynamical equations. According to the Newtonian mechanics, the small compact object moves on the IMBH's equatorial plane, and the angular momentum is

\begin{equation}
L=\mu r^{2}\dot{\phi} ,
\label{L1}
\end{equation}

where $\mu$ is the mass of the small compact object, $r$ is the distance of the binary system, and $\phi$ is the angular position of the small compact object. Based on the equation (\ref{L1}), the total energy is written as

\begin{equation}
E=\dfrac{1}{2}\mu\dot{r}^{2}+\dfrac{1}{2}\mu r^{2}\dot{\phi}^{2}-\dfrac{GM\mu}{r}=\dfrac{1}{2}\mu\dot{r}^{2}+\dfrac{L^{2}}{2\mu r^{2}}-\dfrac{GM\mu}{r} .
\label{E1}
\end{equation}
Here without considering the dissipation, the energy $E$ and the angular momentum $L$ are conserved. The semi-latus rectum $p$ and the eccentricity $e$ can be described by $E$ and $L$ \citep{2007}, and the results are
   
\begin{equation}
p=\dfrac{L^{2}}{GM\mu^{2}}
\label{p_L}
\end{equation}
and
\begin{equation}
e^{2}=1+\dfrac{2EL^{2}}{G^{2}M^{2}\mu^{3}} .
\label{e2}
\end{equation}

We now consider the effects of GW emission and dynamical friction into the binary system, and then the $E$ and $L$ will no longer be conserved. Differentiating the equations (\ref{p_L}) and (\ref{e2}), we can obtain
\begin{equation}
\dot{p}=\dfrac{2L}{GM\mu^{2}}\dot{L}=\dfrac{2}{\mu}\sqrt{\dfrac{p}{GM}}\dot{L}
\label{p_t}
\end{equation}
and
\begin{equation}
\dot{e}=\dfrac{p}{GM\mu e}\dot{E}-\dfrac{(1-e^{2})}{e\mu\sqrt{GMp}}\dot{L} .
\label{e_t}
\end{equation}

In the case of adiabatic approximation, we consider $\dot{E}$ and $\dot{L}$ as the time-averaged rates, and they are
\begin{equation}
\dot{E}=\langle\dfrac{dE}{dt}\rangle_{\rm{GW}}+\langle\dfrac{dE}{dt}\rangle_{\rm{DF}} ,
\label{Et}
\end{equation}

\begin{equation}
\dot{L}=\langle\dfrac{dL}{dt}\rangle_{\rm{GW}}+\langle\dfrac{dL}{dt}\rangle_{\rm{DF}} , 
\label{Lt}
\end{equation}
where the symbol $\langle\rangle$ represents the time average, the subscripts GW represents the loss of the energy caused by GW emission, and the subscripts DF represents the loss of the angular momentum caused by dynamical friction.

Based on the minimum order of post-Newtonian approximation \citep{2007, 1964PhysRev..136.B1224, 1963PhysRev..131.435}, the loss of energy and angular momentum due to GW can be given by
\begin{equation}
\langle\dfrac{dE}{dt}\rangle_{\rm{GW}}=-\dfrac{32}{5}\dfrac{G^{4}\mu^{2}M^{3}}{c^{5}p^{5}}(1-e^{2})^{\dfrac{3}{2}}(1+\dfrac{73}{24}e^{2}+\dfrac{37}{96}e^{4}) ,
\label{Et_GW}
\end{equation}

\begin{equation}
\langle\dfrac{dL}{dt}\rangle_{\rm{GW}}=-\dfrac{32}{5}\dfrac{G^{\dfrac{7}{2}}\mu^{2}M^{\dfrac{5}{2}}}{c^{5}p^{\dfrac{7}{2}}}(1-e^{2})^{\dfrac{3}{2}}(1+\dfrac{7}{8}e^{2}) .
\label{Lt_GW}
\end{equation}

When the stellar massive compact object passes through the DM halo around the IMBH, it gravitationally interacts with the DM particles, this effect is called dynamical friction or gravitational drag \citep{1943ApJ....97..255C}. The dynamical friction force is given by \citep{2015PhRvD..91d4045E, 1943ApJ....97..255C} 
\begin{equation}
F_{\rm{DF}}=\dfrac{4\pi G^{2}\mu^{2}\rho_{\rm{DM}}(r)ln\Lambda}{\upsilon^{2}} .
\label{F}
\end{equation}

Here we set the Coulomb logarithm $ln\Lambda\simeq 10$ \citep{2007CQGra..24R.113A}. The radius $r$ satisfies $r=p/(1+e\cos\phi)$. The total energy $E$ is the sum of the gravitational potential and the kinetic energy of the stellar massive compact object, i.e. $E=-GM\mu/r+\mu \upsilon^{2}/2$. Then combining equations (\ref{p_L}) and (\ref{e2}), we can get the velocity $\upsilon$ of the small compact object
\begin{equation}
\upsilon=\sqrt{\dfrac{2E}{\mu}+\dfrac{2GM}{r}}=(\dfrac{GM}{p})^{\dfrac{1}{2}}\sqrt{(e^{2}-1)+2(1+e\cos\phi)} .
\label{v}
\end{equation}

According to equations (\ref{New_DM_profile}), (\ref{F}) and (\ref{v}), we can obtain the average energy loss rate caused by dynamical friction
\begin{equation}
\langle\frac{dE}{dt}\rangle_{\rm{DF}}=\dfrac{1}{T}\int_{0}^{T}\dfrac{dE}{dt}\mid_{\rm{DF}}dt=\dfrac{1}{T}\int_{0}^{T}F_{\rm{DF}} \upsilon dt=\dfrac{1}{T}\int_{0}^{T}\dfrac{4\pi G^{2}\mu^{2}\rho_{\rm{DM}}(r)ln\Lambda}{\upsilon} dt  $$$$
=\dfrac{1}{T}\int_{0}^{T}\dfrac{4\pi G^{\dfrac{3}{2}}\mu^{2}\rho_{\rm{sp}}r_{\rm{sp}}^{\alpha}ln\Lambda (1+e\cos\phi)^{\alpha}(p-\dfrac{kGM}{c^{2}}(1+e\cos\phi))^{3}}{p^{\alpha+\dfrac{5}{2}}M^{\dfrac{1}{2}}(1+2e\cos\phi+e^{2})^{\dfrac{1}{2}}} dt  $$$$
=(1-e^{2})^{\dfrac{3}{2}}\int_{0}^{2\pi}\dfrac{2G^{\dfrac{3}{2}}\mu^{2}\rho_{\rm{sp}}r_{\rm{sp}}^{\alpha}ln\Lambda (1+e\cos\phi)^{\alpha-2}(p-\dfrac{kGM}{c^{2}}(1+e\cos\phi))^{3}}{p^{\alpha+\dfrac{5}{2}}M^{\dfrac{1}{2}}(1+2e\cos\phi+e^{2})^{\dfrac{1}{2}}} d\phi .
\label{DF_energy loss rate}
\end{equation}

Based on the geometrical relation, the angular momentum loss rate resulted from dynamical friction can be expressed as $(dL/dt)_{\rm{DF}}=r\cdot F_{\rm{DF}}(r\dot{\phi}/\upsilon)$. Then substituting equations (\ref{New_DM_profile}), (\ref{L1}), (\ref{p_L}), (\ref{F}) and (\ref{v}), and the $r$ is the same as before, so we can get the average loss rate of angular momentum
\begin{equation}
\langle\frac{dL}{dt}\rangle_{\rm{DF}}=\dfrac{1}{T}\int_{0}^{T}\dfrac{dL}{dt}\mid_{\rm{DF}}dt=\dfrac{1}{T}\int_{0}^{T}\dfrac{4\pi G\mu^{2}\rho_{\rm{DM}}(r)p^{2}ln\Lambda}{M(1+2e\cos\phi+e^{2})^{\dfrac{3}{2}}}dt   $$$$
=\dfrac{1}{T}\int_{0}^{T}\dfrac{4\pi G\mu^{2}\rho_{\rm{sp}}r_{\rm{sp}}^{\alpha}ln\Lambda(1+e\cos\phi)^{\alpha}(p-\dfrac{kGM}{c^{2}}(1+e\cos\phi))^{3}}{p^{\alpha+1}M(1+2e\cos\phi+e^{2})^{\dfrac{3}{2}}}dt  $$$$
=(1-e^{2})^{\dfrac{3}{2}}\int_{0}^{2\pi}\dfrac{2G\mu^{2}\rho_{\rm{sp}}r_{\rm{sp}}^{\alpha}ln\Lambda(1+e\cos\phi)^{\alpha-2}(p-\dfrac{kGM}{c^{2}}(1+e\cos\phi))^{3}}{p^{\alpha+1}M(1+2e\cos\phi+e^{2})^{\dfrac{3}{2}}}d\phi .
\label{DF_angular momentum loss rate}
\end{equation}

In the above equations (\ref{DF_energy loss rate}) and (\ref{DF_angular momentum loss rate}), we have used a relation in the last step, i.e. $ \int_{0}^{T}(...)\dfrac{dt}{T}=(1-e^{2})^{\dfrac{3}{2}}\int_{0}^{2\pi}(1+e\cos\phi)^{-2}(...)\dfrac{d\phi}{2\pi} $. Substituting equations (\ref{Et_GW}), (\ref{Lt_GW}), (\ref{DF_energy loss rate}) and (\ref{DF_angular momentum loss rate}) into (\ref{Et}) and (\ref{Lt}), we can obtain the total loss rate of energy and angular momentum resulted from GW and DF

\begin{equation}
\dot{E}=-\dfrac{32}{5}\dfrac{G^{4}\mu^{2}M^{3}}{c^{5}p^{5}}(1-e^{2})^{\dfrac{3}{2}}(1+\dfrac{73}{24}e^{2}+\dfrac{37}{96}e^{4})-\dfrac{2G^{\dfrac{3}{2}}\mu^{2}\rho_{\rm{sp}}r_{\rm{sp}}^{\alpha}ln\Lambda}{p^{\alpha+\dfrac{5}{2}}M^{\dfrac{1}{2}}}(1-e^{2})^{\dfrac{3}{2}}\int_{0}^{2\pi}\dfrac{(1+e\cos\phi)^{\alpha-2}(p-\dfrac{kGM}{c^{2}}(1+e\cos\phi))^{3}}{(1+2e\cos\phi+e^{2})^{\dfrac{1}{2}}}d\phi ,
\label{total energy loss rate}
\end{equation}

\begin{equation}
\dot{L}=-\dfrac{32}{5}\dfrac{G^{\dfrac{7}{2}}\mu^{2}M^{\dfrac{5}{2}}}{c^{5}p^{\dfrac{7}{2}}}(1-e^{2})^{\dfrac{3}{2}}(1+\dfrac{7}{8}e^{2})-\dfrac{2G\mu^{2}\rho_{\rm{sp}}r_{\rm{sp}}^{\alpha}ln\Lambda}{p^{\alpha+1}M}(1-e^{2})^{\dfrac{3}{2}}\int_{0}^{2\pi}\dfrac{(1+e\cos\phi)^{\alpha-2}(p-\dfrac{kGM}{c^{2}}(1+e\cos\phi))^{3}}{(1+2e\cos\phi+e^{2})^{\dfrac{3}{2}}}d\phi .
\label{total angular momentum loss rate}
\end{equation}

Substituting equations (\ref{total energy loss rate}) and (\ref{total angular momentum loss rate}) into (\ref{p_t}) and (\ref{e_t}), we can obtain the dynamical equations of IMRI under the effect of a DM spike
\begin{equation}
\dot{p}=-\dfrac{64}{5}\dfrac{G^{3}\mu M^{2}}{c^{5}p^{3}}(1-e^{2})^{\dfrac{3}{2}}(1+\dfrac{7}{8}e^{2})-\dfrac{4G^{\dfrac{1}{2}}\mu \rho_{\rm{sp}}r_{\rm{sp}}^{\alpha}ln\Lambda}{p^{\alpha+\dfrac{1}{2}}M^{\dfrac{3}{2}}}(1-e^{2})^{\dfrac{3}{2}}\int_{0}^{2\pi}\dfrac{(1+e\cos\phi)^{\alpha-2}(p-\dfrac{kGM}{c^{2}}(1+e\cos\phi))^{3}}{(1+2e\cos\phi+e^{2})^{\dfrac{3}{2}}}d\phi ,
\label{new p_t}
\end{equation}

\begin{equation}
\dot{e}=-\dfrac{304}{15}\dfrac{G^{3}\mu M^{2}}{c^{5}p^{4}}(1-e^{2})^{\dfrac{3}{2}}e(1+\dfrac{121}{304}e^{2})$$$$
-\dfrac{4G^{\dfrac{1}{2}}\mu \rho_{\rm{sp}}r_{\rm{sp}}^{\alpha}ln\Lambda}{p^{\alpha+\dfrac{3}{2}}M^{\dfrac{3}{2}}}(1-e^{2})^{\dfrac{3}{2}}\int_{0}^{2\pi}\dfrac{(e+\cos\phi)(1+e\cos\phi)^{\alpha-2}(p-\dfrac{kGM}{c^{2}}(1+e\cos\phi))^{3}}{(1+2e\cos\phi+e^{2})^{\dfrac{3}{2}}}d\phi ,
\label{new e_t}
\end{equation}

According to the relation of semi-major axis $a$ and semi-latus rectum $p$, i.e. $ a=p/(1-e^{2}) $, we can use $a$ instead of $p$ to describe the dynamical equations of IMRI and get them
\begin{equation}
\dot{e}=-\dfrac{304}{15}\dfrac{G^{3}\mu M^{2}}{c^{5}a^{4}}(1-e^{2})^{-\dfrac{5}{2}}e(1+\dfrac{121}{304}e^{2})$$$$
-\dfrac{4G^{\dfrac{1}{2}}\mu \rho_{\rm{sp}}r_{\rm{sp}}^{\alpha}ln\Lambda}{a^{\alpha+\dfrac{3}{2}}M^{\dfrac{3}{2}}}(1-e^{2})^{-\alpha}\int_{0}^{2\pi}\dfrac{(e+\cos\phi)(1+e\cos\phi)^{\alpha-2}(a(1-e^{2})-\dfrac{kGM}{c^{2}}(1+e\cos\phi))^{3}}{(1+2e\cos\phi+e^{2})^{\dfrac{3}{2}}}d\phi ,
\label{new e_t_a}
\end{equation}

\begin{equation}
\dot{a}=-\dfrac{64}{5}\dfrac{G^{3}\mu M^{2}}{c^{5}a^{3}}(1-e^{2})^{-\dfrac{7}{2}}(1+\dfrac{73}{24}e^{2}+\dfrac{37}{96}e^{4})$$$$
-\dfrac{4G^{\dfrac{1}{2}}\mu \rho_{\rm{sp}}r_{\rm{sp}}^{\alpha}ln\Lambda}{a^{\alpha+\dfrac{1}{2}}M^{\dfrac{3}{2}}}(1-e^{2})^{-\alpha-1}\int_{0}^{2\pi}\dfrac{(1+e\cos\phi)^{\alpha-2}(a(1-e^{2})-\dfrac{kGM}{c^{2}}(1+e\cos\phi))^{3}}{(1+2e\cos\phi+e^{2})^{\dfrac{1}{2}}}d\phi .
\label{a_t}
\end{equation}

\begin{figure*}[htbp]
\centering
\includegraphics[height=5cm,width=7cm]{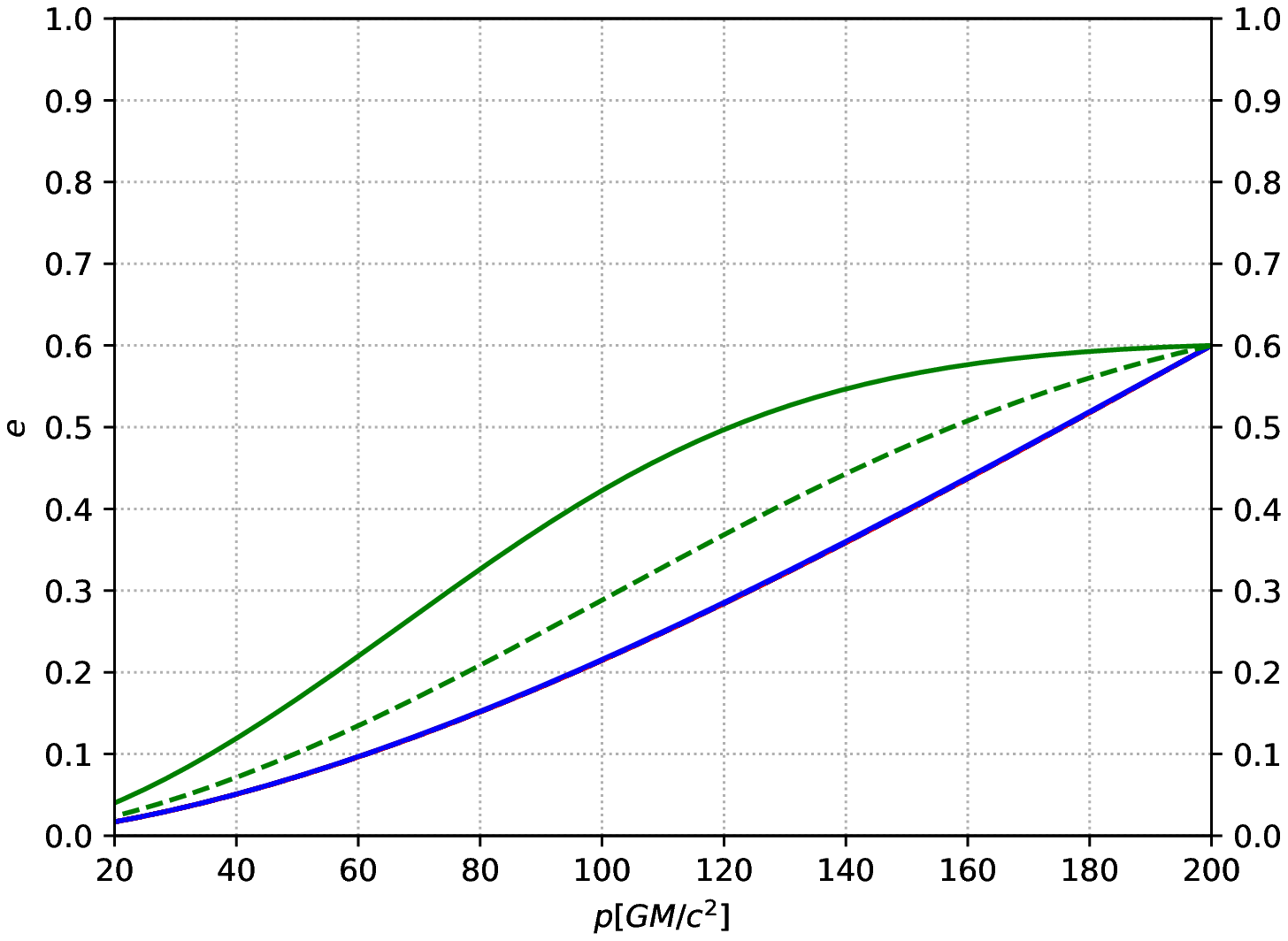}
\includegraphics[height=5cm,width=7cm]{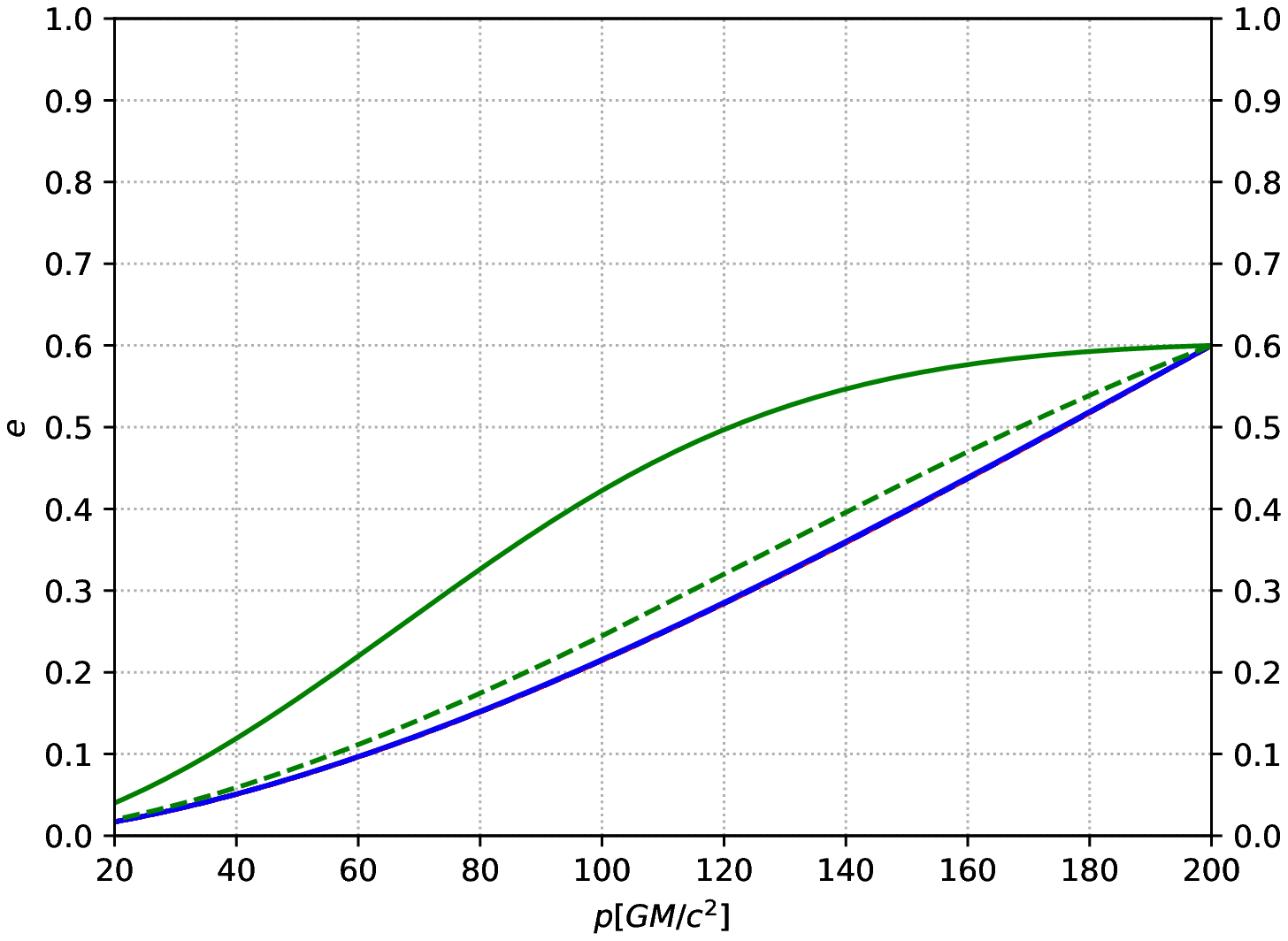}
\includegraphics[height=5cm,width=7cm]{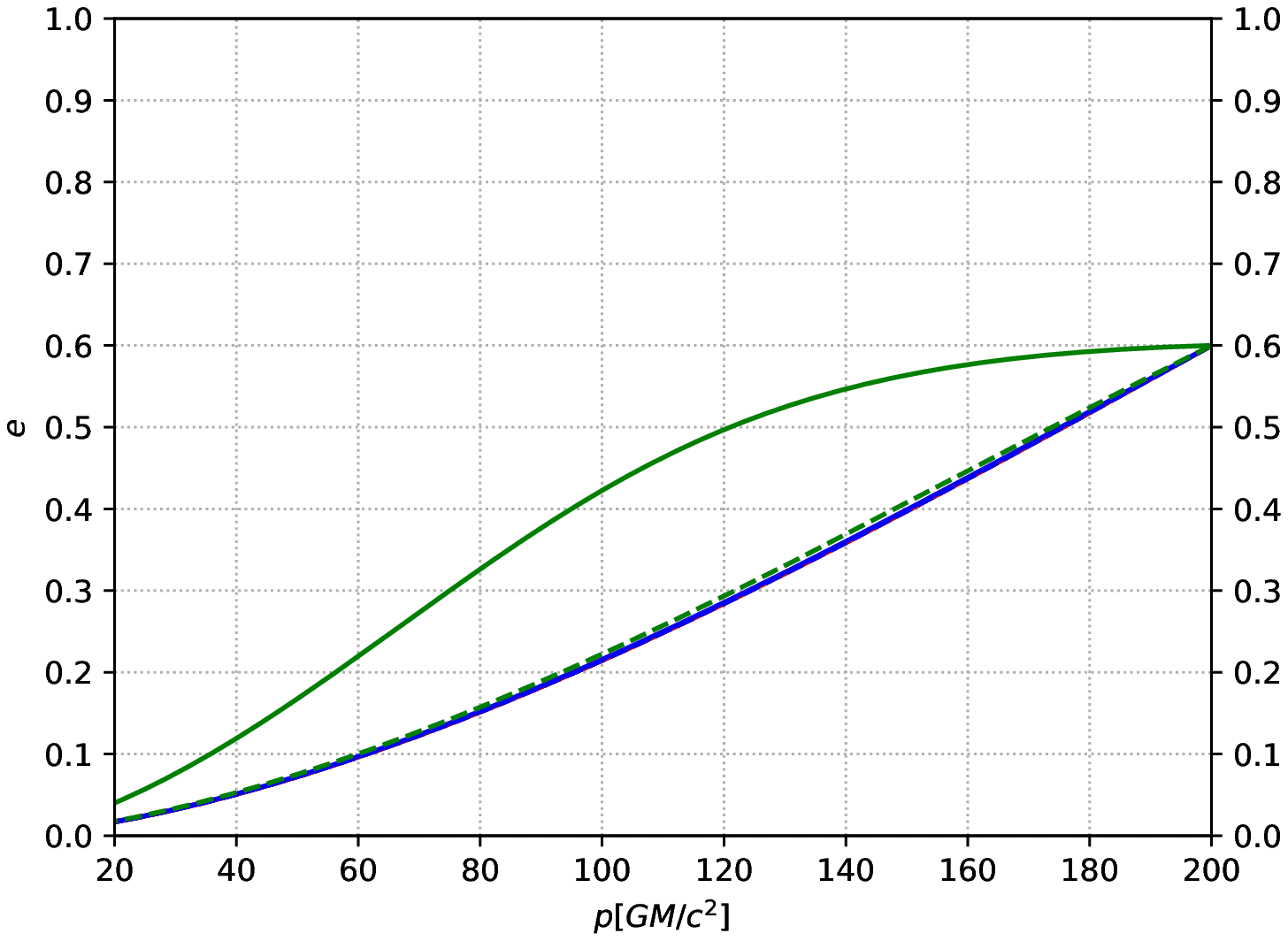}
\includegraphics[height=5cm,width=7cm]{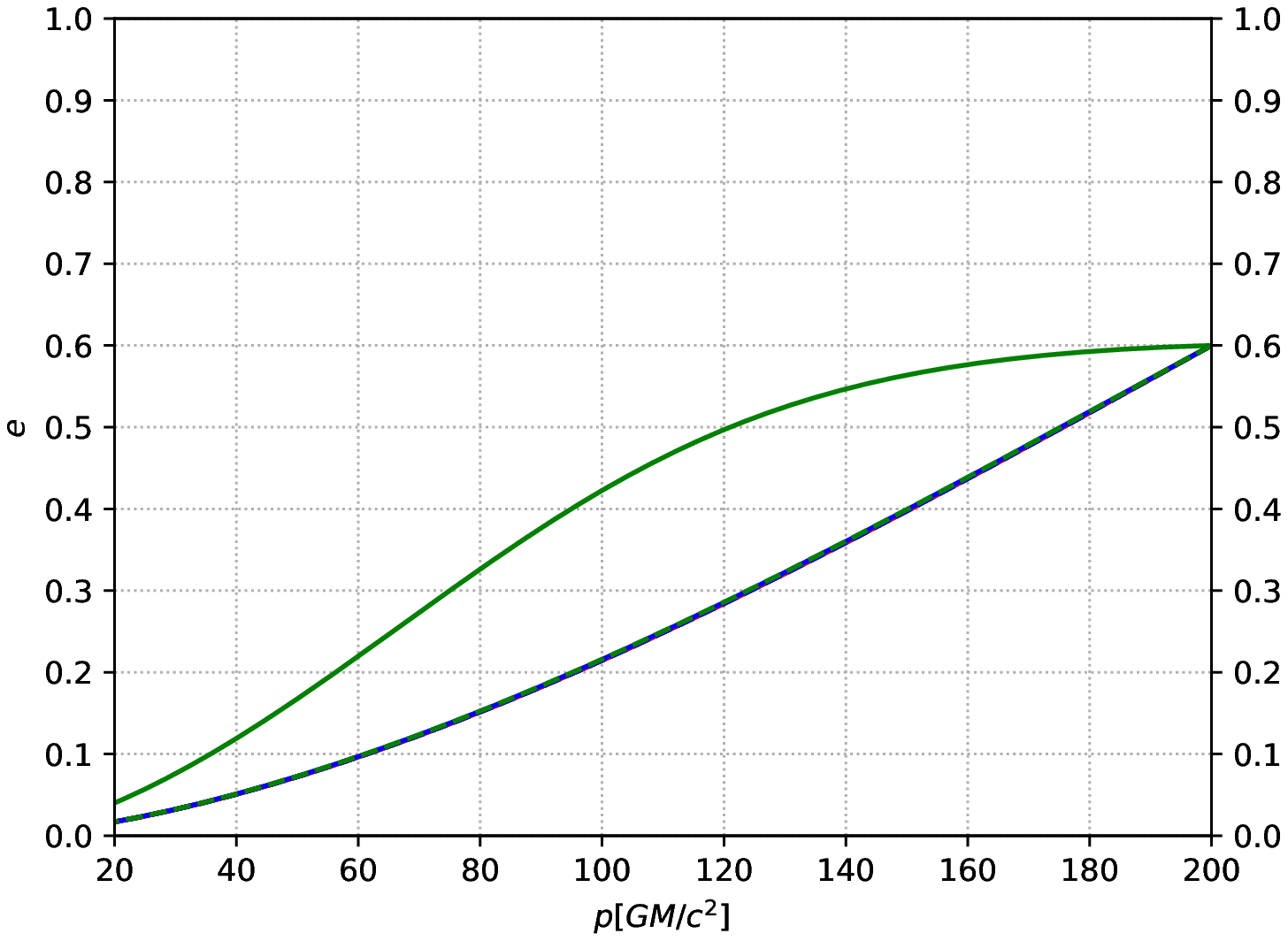}
\caption{The eccentricity $e$ of an IMRI evolves with the semi-latus rectum $p$ under different masses of the central IMBH. The horizontal axis is the semi-latus rectum $p$ with unit of $GM/c^{2}$, the vertical axis is the eccentricity $e$. In this figure, the solid lines represent that the IMBH's mass is $10^{3} M_{\odot}$, and the dashed lines from left to right represent that the IMBH's mass is $1.5\times10^{3} M_{\odot}$, $2\times10^{3} M_{\odot}$, $3\times10^{3} M_{\odot}$ and $5\times10^{3} M_{\odot}$, respectively. We take the small compact object's mass as $10 M_{\odot}$ and the initial $p$ as $200 GM/c^{2}$. The black lines correspond to the absence of DM, and the red, blue and green lines correspond to $\alpha=1.5$, $2.0$ and $7/3$, respectively. }
\label{result1}
\end{figure*}

\begin{figure*}[htbp]
\centering
\includegraphics[height=5cm,width=7cm]{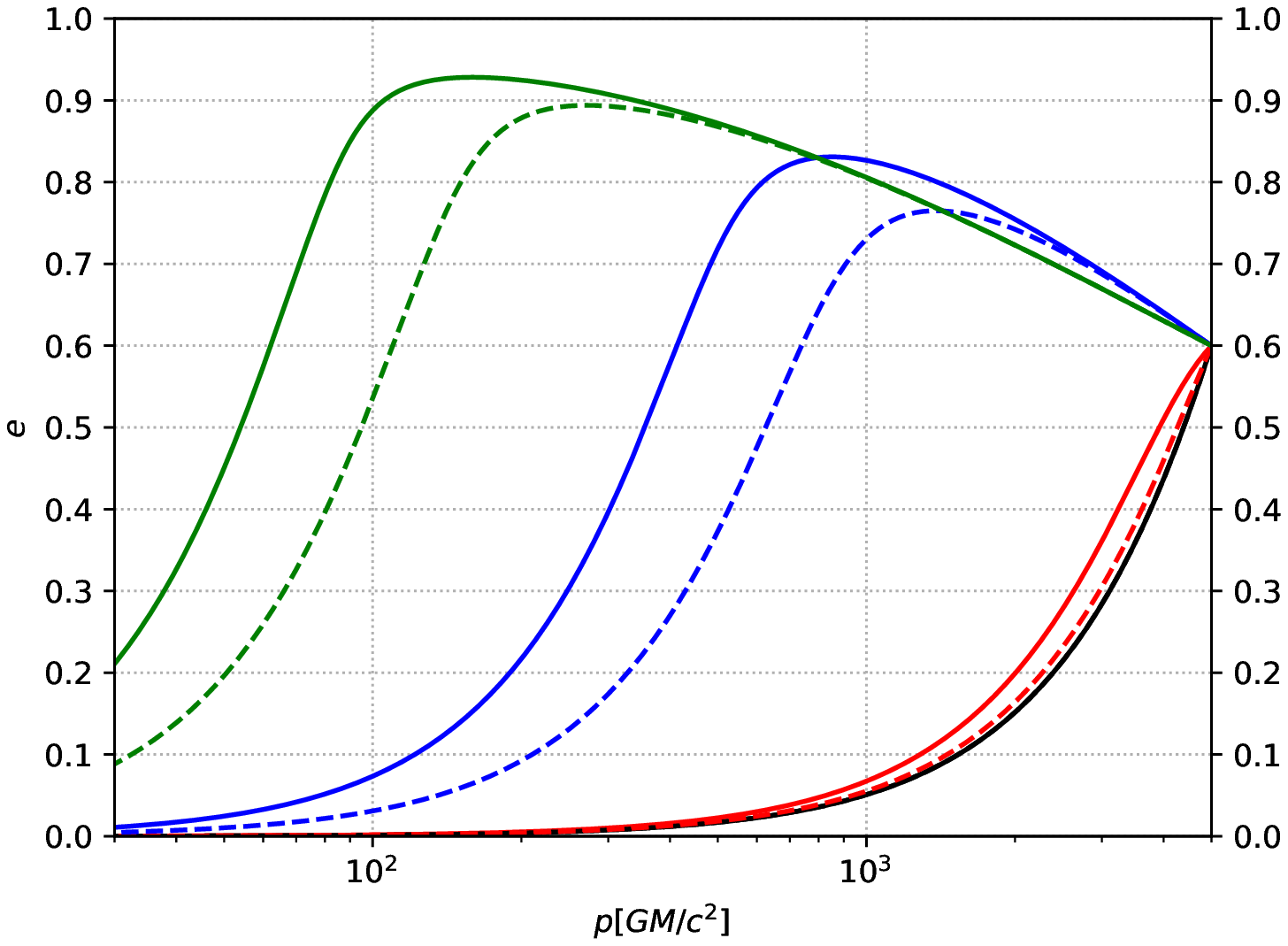}
\includegraphics[height=5cm,width=7cm]{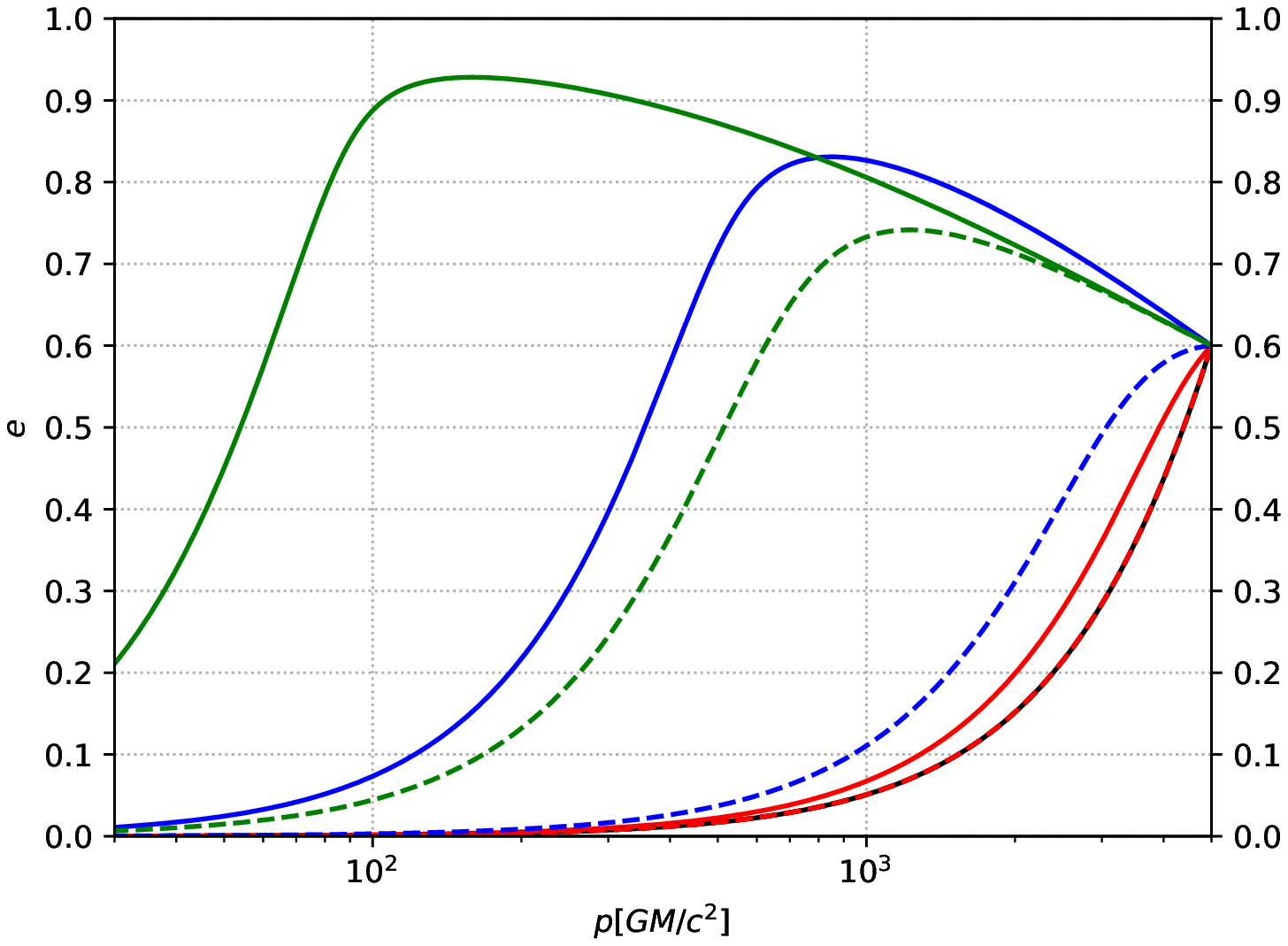}
\includegraphics[height=5cm,width=7cm]{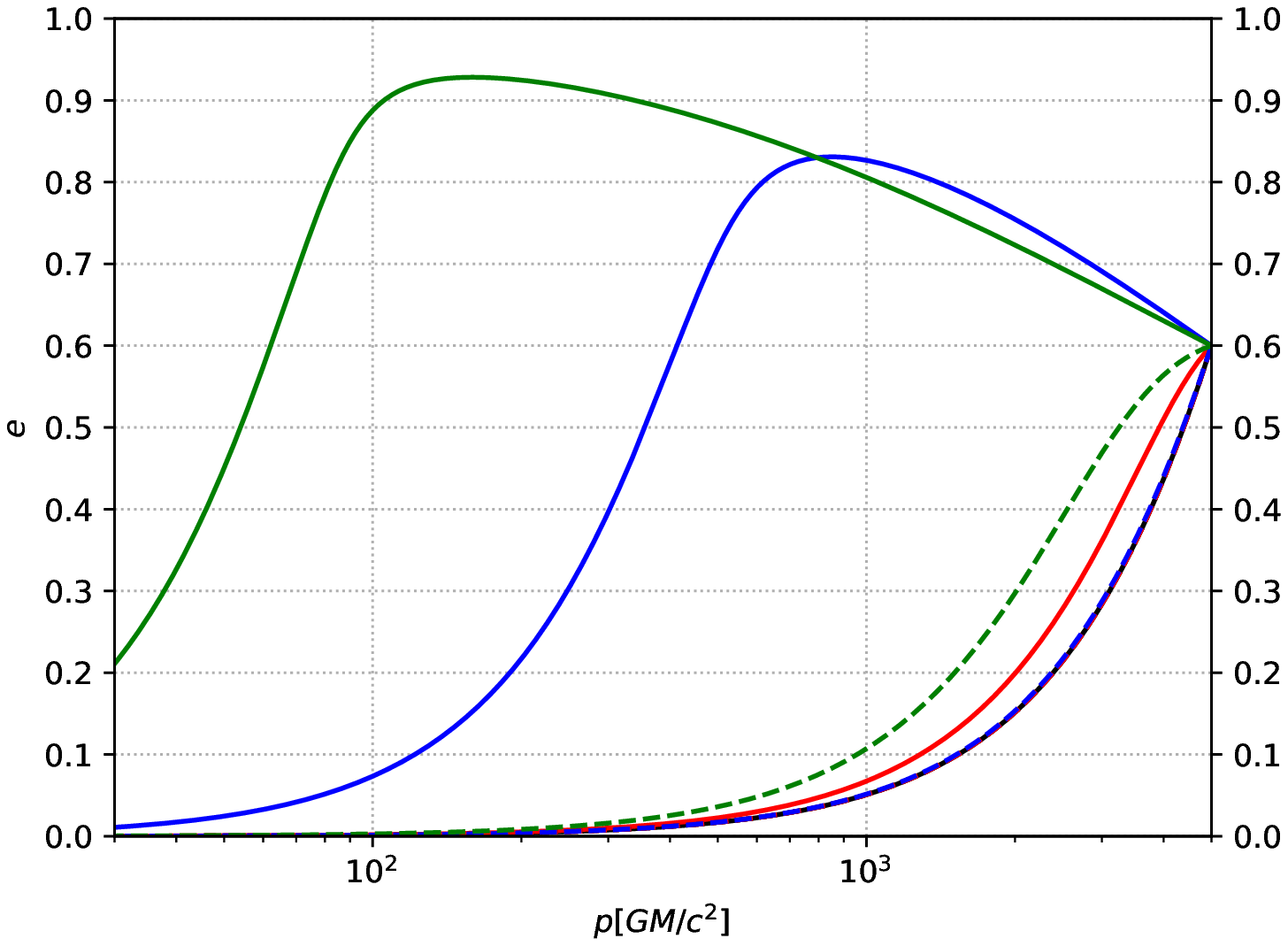}
\includegraphics[height=5cm,width=7cm]{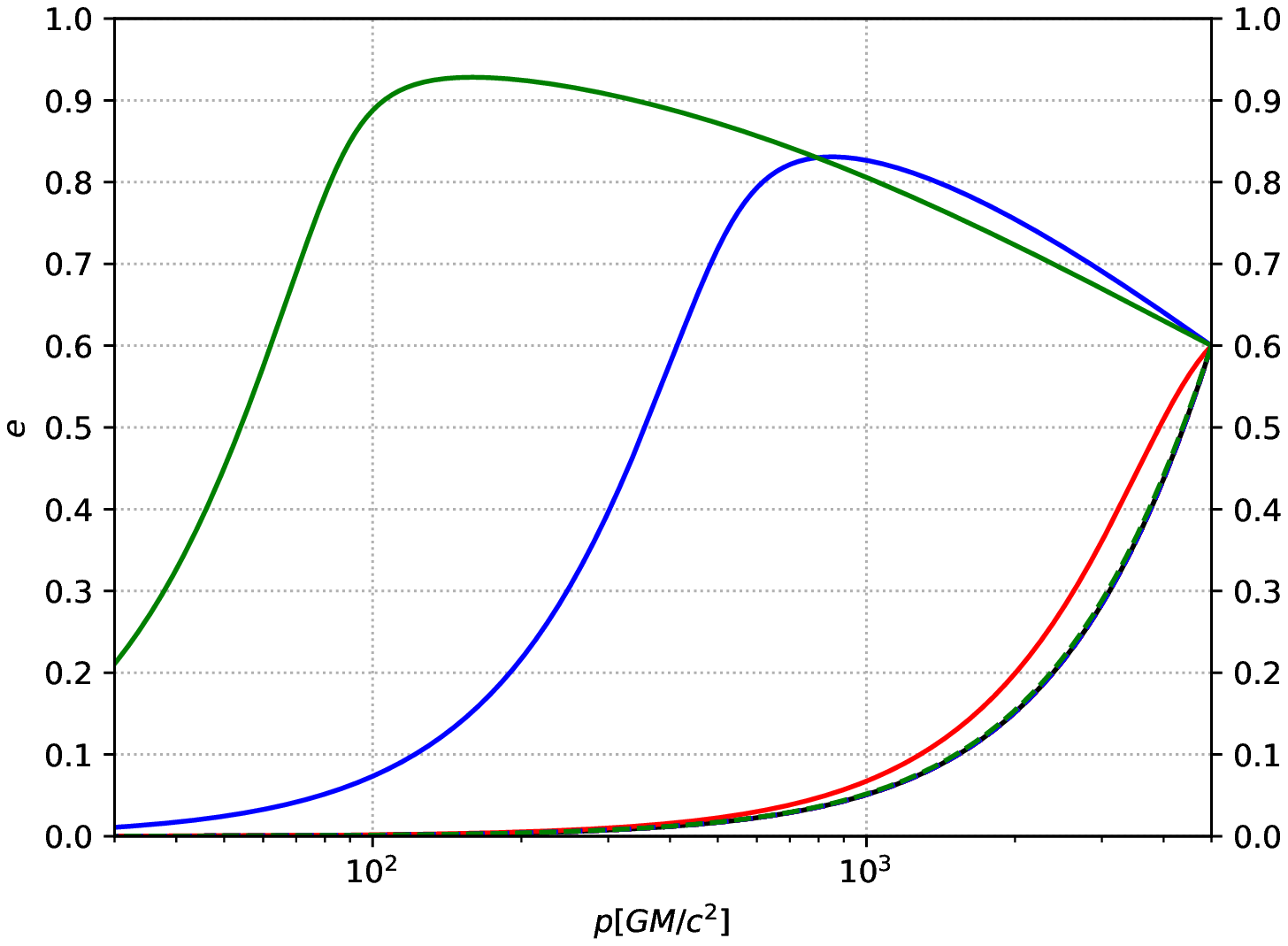}
\caption{The eccentricity $e$ of an IMRI evolves with the semi-latus rectum $p$ under different masses of the central IMBH. The horizontal axis is the semi-latus rectum $p$ with unit of $GM/c^{2}$, the vertical axis is the eccentricity $e$. In this figure, the solid lines represent that the IMBH's mass is $10^{3} M_{\odot}$, and the dashed lines from left to right represent that the IMBH's mass is $1.5\times10^{3} M_{\odot}$, $5\times10^{3} M_{\odot}$, $2\times10^{4} M_{\odot}$ and $7\times10^{4} M_{\odot}$, respectively. We take the small compact object's mass as $10 M_{\odot}$ and the initial $p$ as $5000 GM/c^{2}$. The black lines correspond to the absence of DM, and the red, blue and green lines correspond to $\alpha=1.5$, $2.0$ and $7/3$, respectively. }
\label{result11}
\end{figure*}

\begin{figure*}[htbp]
\centering
\includegraphics[height=5cm,width=7cm]{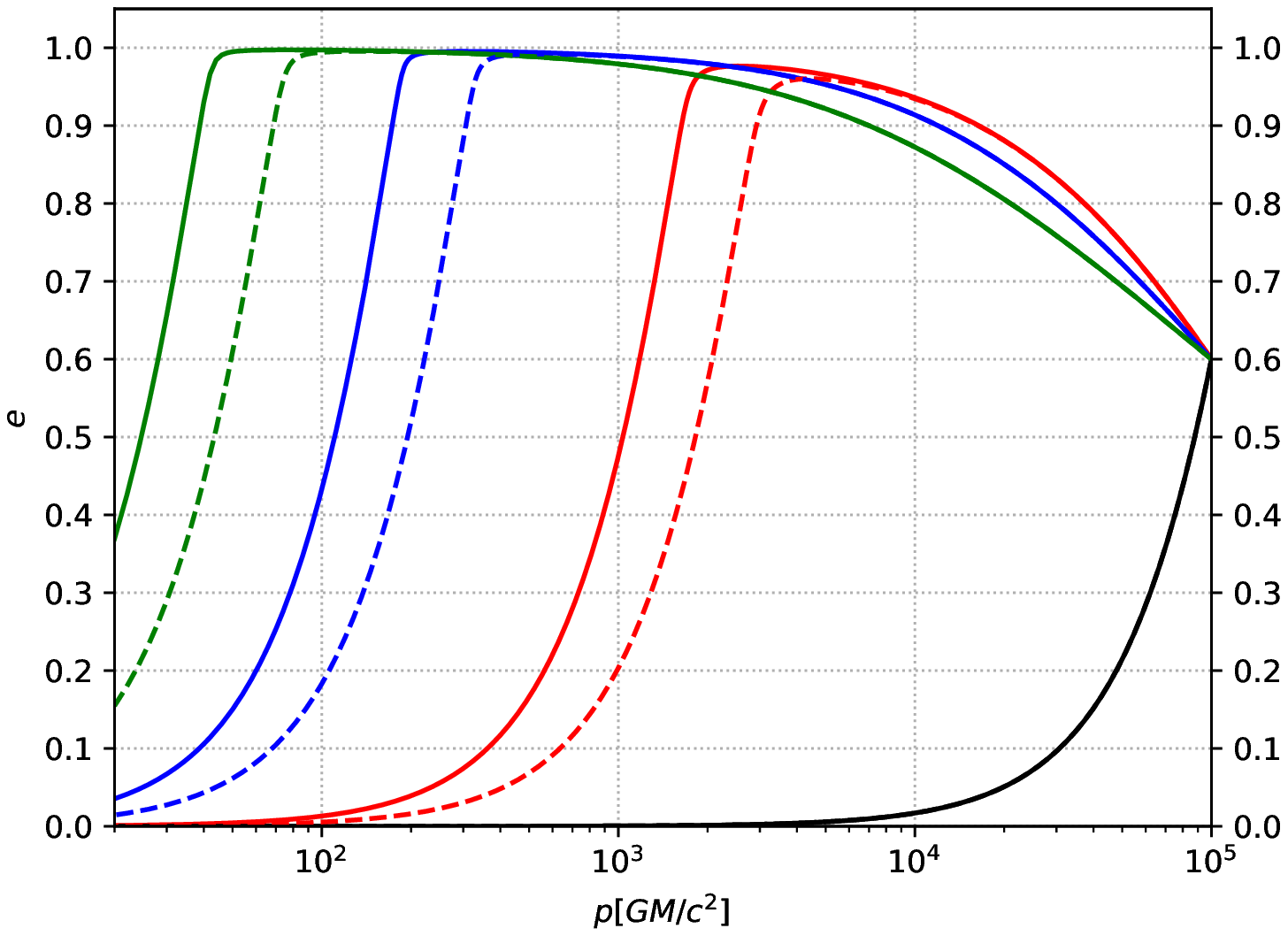}
\includegraphics[height=5cm,width=7cm]{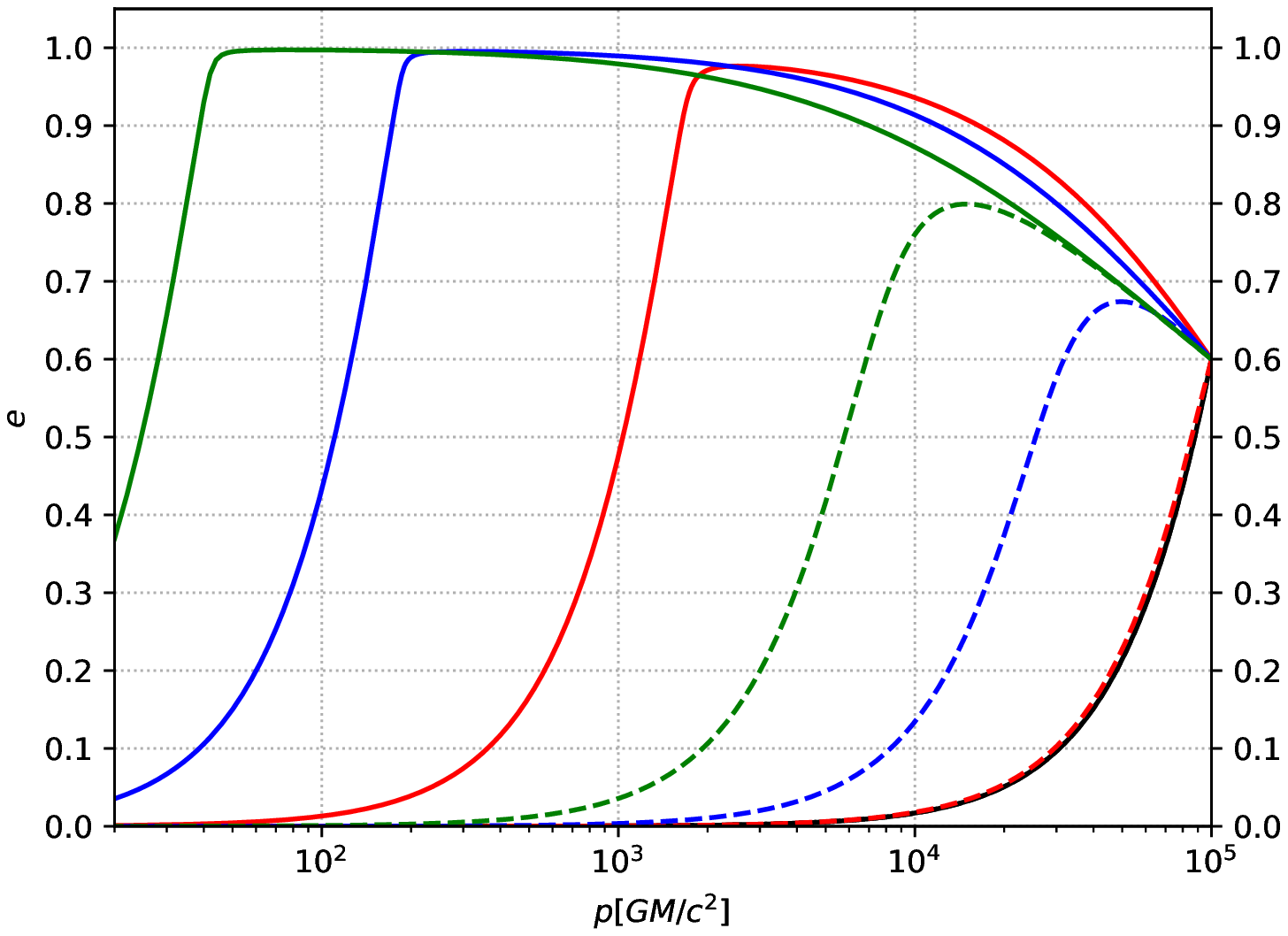}
\includegraphics[height=5cm,width=7cm]{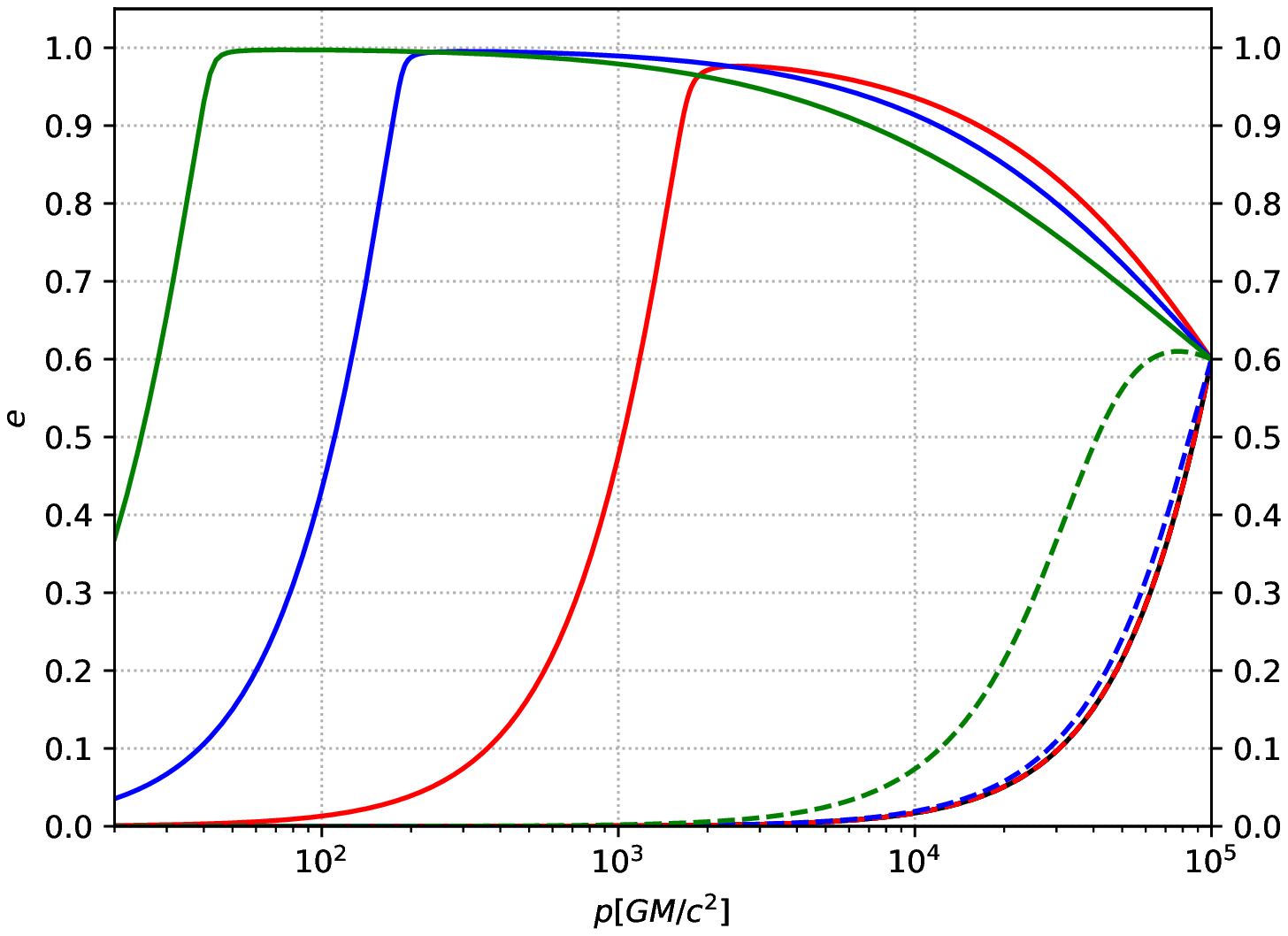}
\includegraphics[height=5cm,width=7cm]{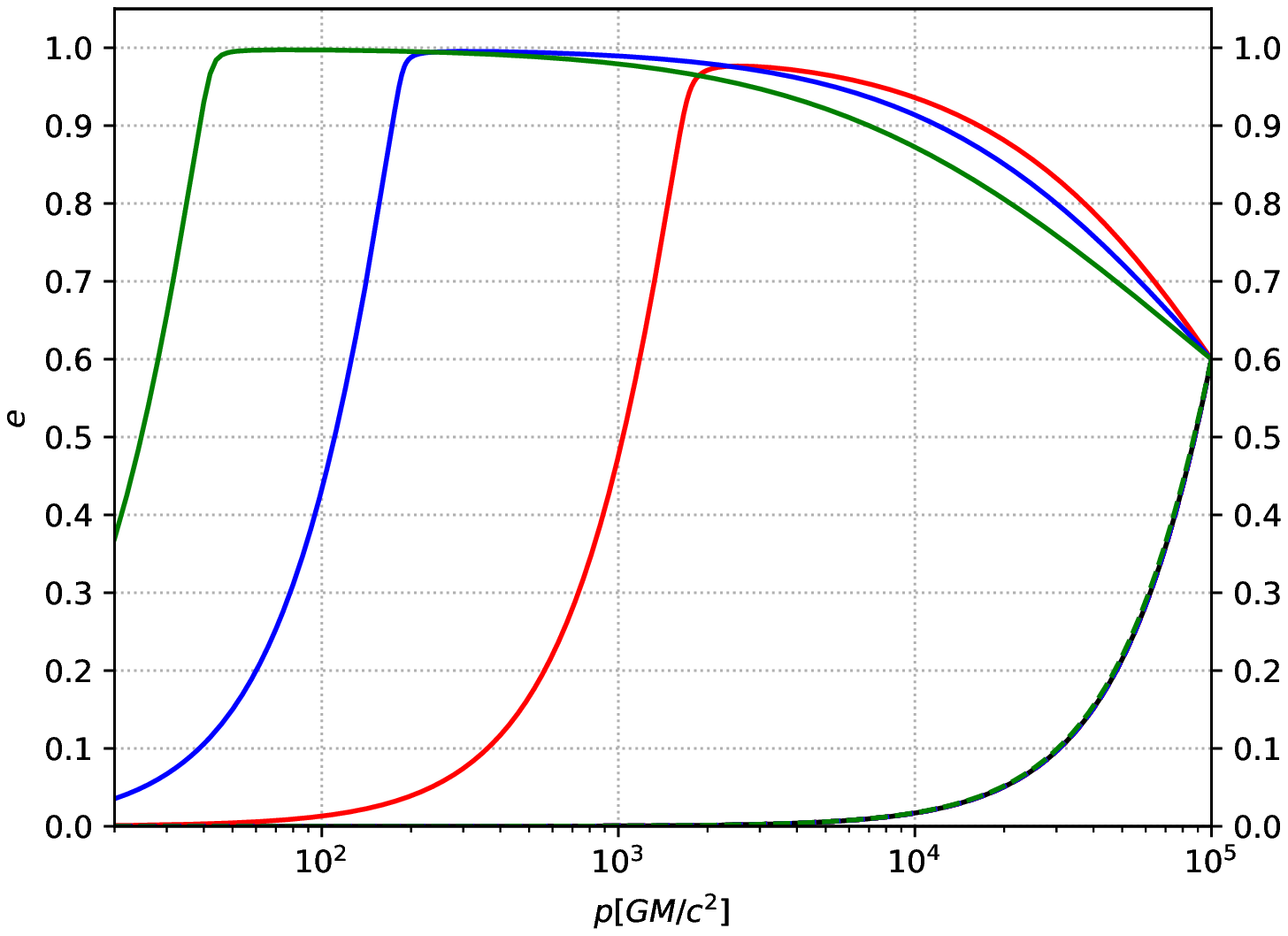}
\caption{The eccentricity $e$ of an IMRI evolves with the semi-latus rectum $p$ under different masses of the central IMBH. The horizontal axis is the semi-latus rectum $p$ with unit of $GM/c^{2}$, the vertical axis is the eccentricity $e$. In this figure, the solid lines represent that the IMBH's mass is $10^{3} M_{\odot}$, and the dashed lines from left to right represent that the IMBH's mass is $1.5\times10^{3} M_{\odot}$, $5\times10^{4} M_{\odot}$, $2\times10^{5} M_{\odot}$ and $10^{6} M_{\odot}$, respectively. We take the small compact object's mass as $10 M_{\odot}$ and the initial $p$ as $10^{5} GM/c^{2}$. The black lines correspond to the absence of DM, and the red, blue and green lines correspond to $\alpha=1.5$, $2.0$ and $7/3$, respectively. }
\label{result111}
\end{figure*}

Figure \ref{result1}, \ref{result11} and \ref{result111}  describe the p-e relation under different initial conditions, different profiles of DM spikes and different masses of the center IMBH. When the initial $p$ is relatively small, as shown in Figure \ref{result1}, for different IMBH's masses, the curves of $\alpha=1.5$ and $2.0$ are basically the same as the cases without DM. For $\alpha=7/3$, the relatively denser DM spike can still decrease the orbit circularization rate of an IMRI. As the mass of the center IMBH increases, the curves for $\alpha=7/3$ gradually tend to cases without DM. When the initial $p$ is relatively large, as shown in Figure \ref{result11} and \ref{result111}, for $\alpha=1.5$, $2.0$ and $7/3$, the DM spike can significantly increase the eccentricity. In some cases, the eccentricity can even close to $1$. When the eccentricity is close to $1$, part of the three curves for $\alpha=1.5$, $2.0$ and $7/3$ will overlap, as shown in Figure \ref{result111}. As the mass of the IMBH increases, the three curves gradually tend to the cases without DM.

\begin{figure*}[htbp]
\centering
\includegraphics[height=5cm,width=7cm]{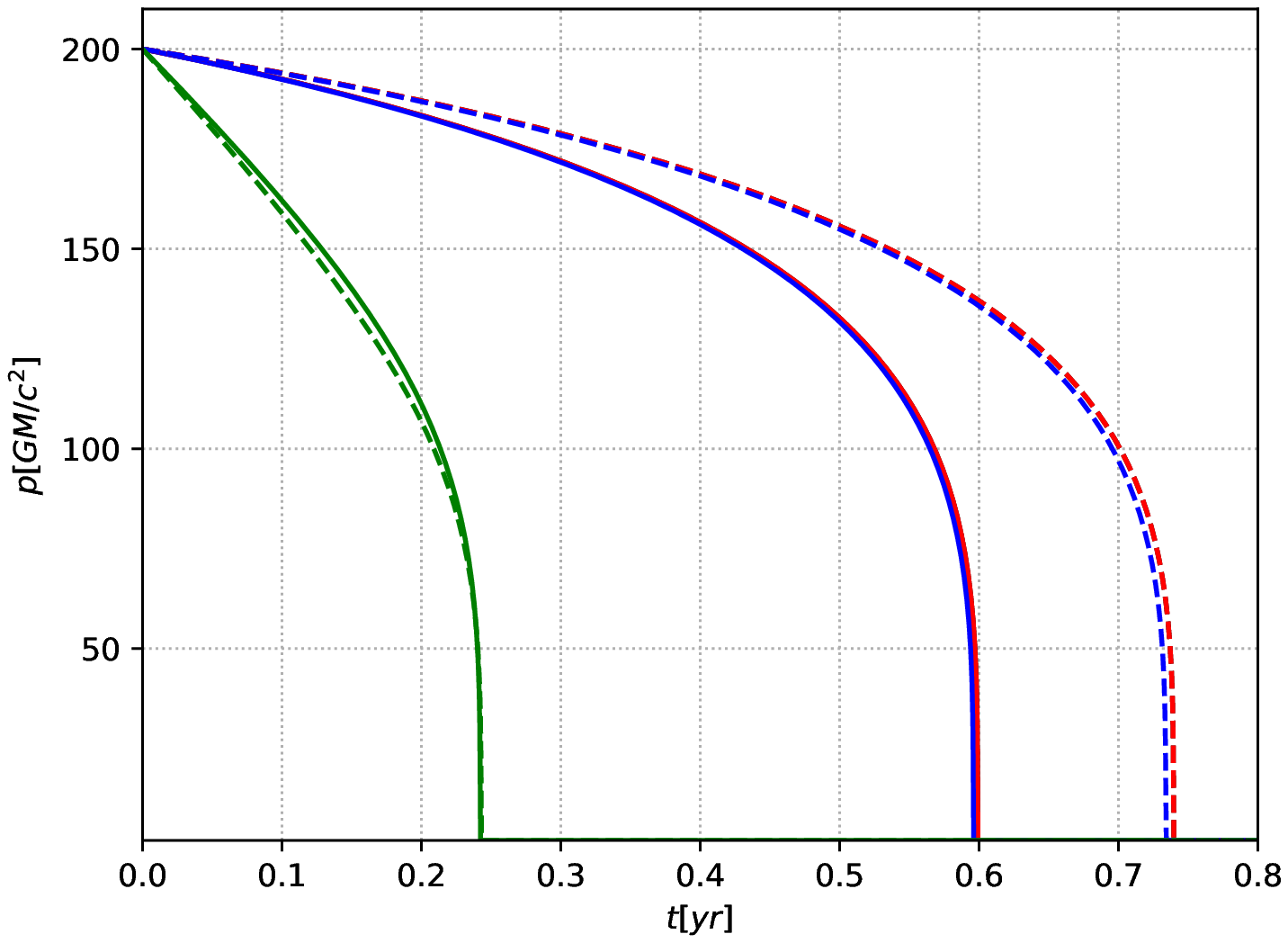}
\includegraphics[height=5cm,width=7cm]{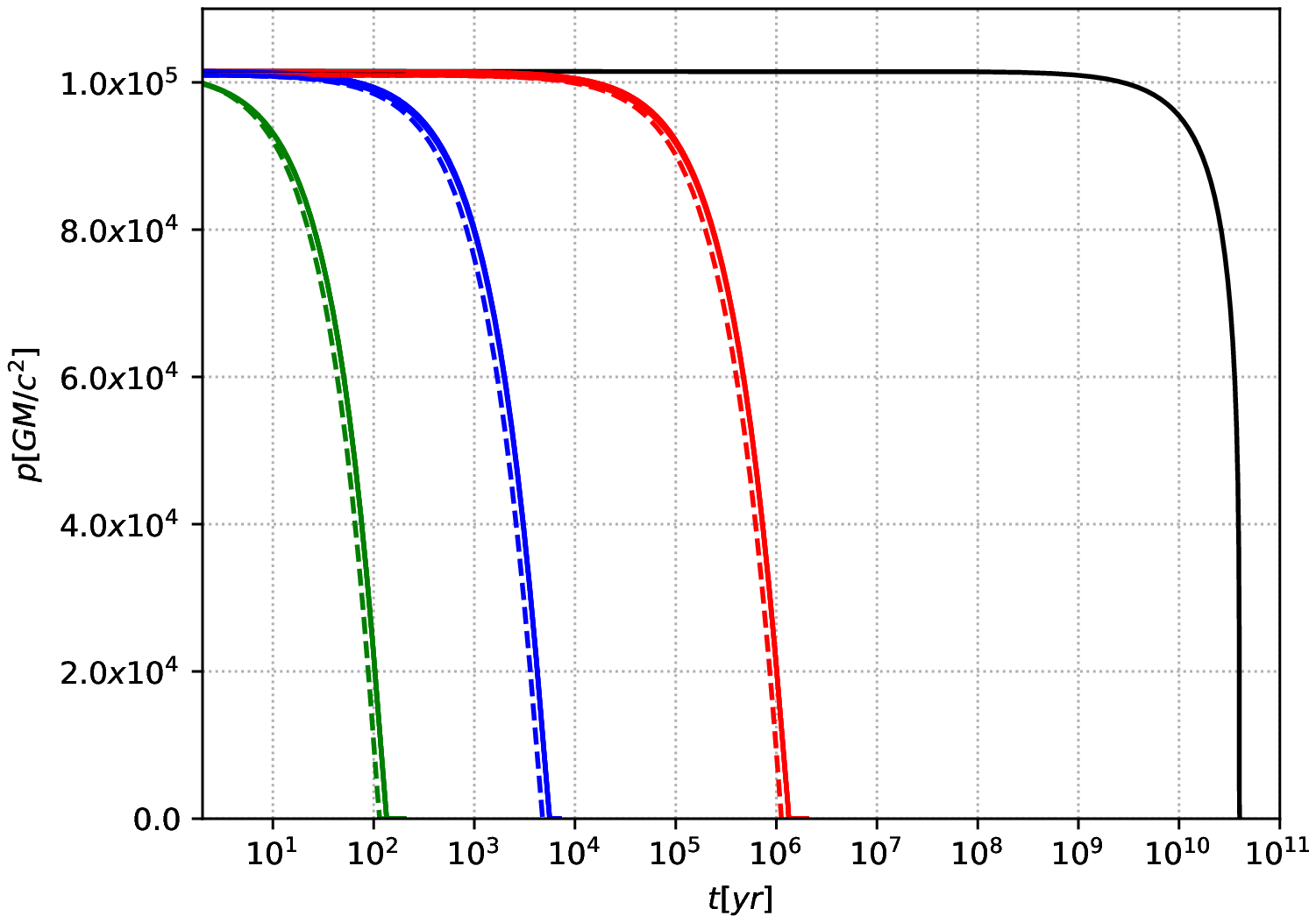}
\includegraphics[height=5cm,width=7cm]{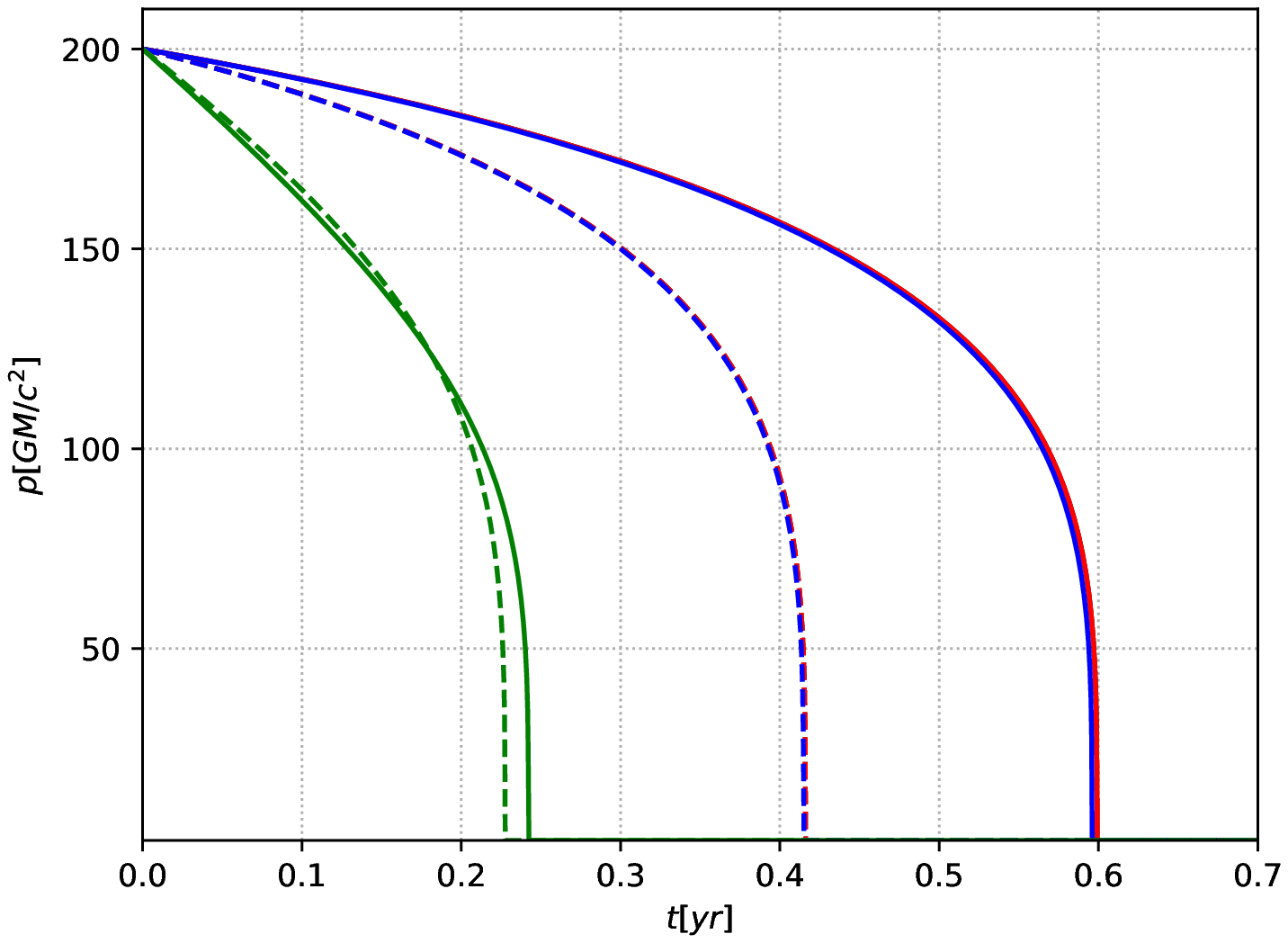}
\includegraphics[height=5cm,width=7cm]{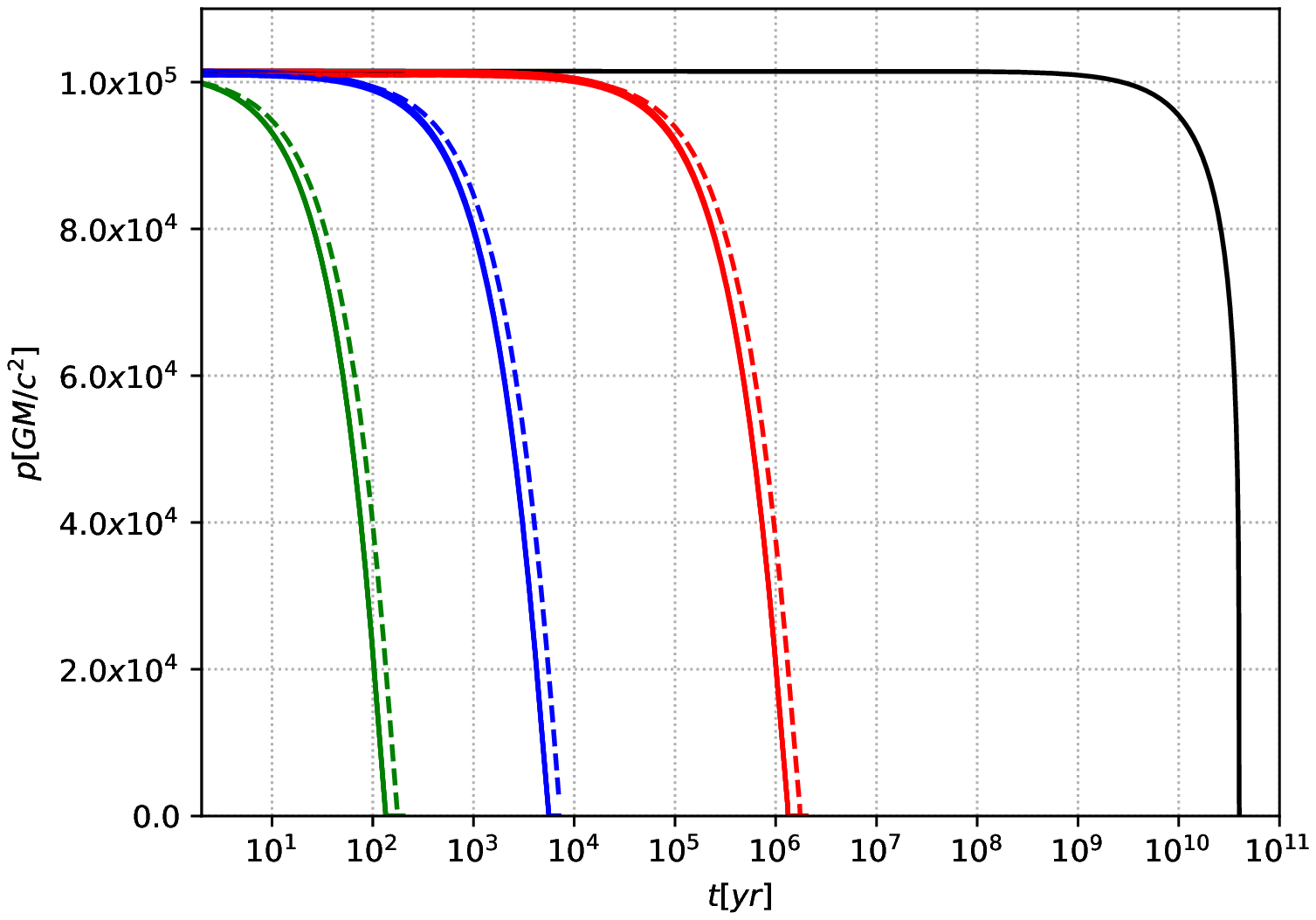}
\caption{The semi-latus rectum $p$ of an IMRI evolves with time $t$ under different IMBH's masses. The horizontal axis is time $t$ with unit of year (yr), the vertical axis is the semi-latus rectum$p$ with unit of $GM/c^{2}$. We take the small compact object's mass as $10 M_{\odot}$ and the initial eccentricity $e=0.6$. The solid lines represent that the IMBH's mass is $10^{3} M_{\odot}$. The dashed lines represent that the IMBH's mass is $9\times10^{2} M_{\odot}$ for the upper panels and $1.2\times10^{3} M_{\odot}$ for the lower panels. The left panels: the initial $p$ is set to $200 GM/c^{2}$. The right panels: the relatively large initial is $p\simeq10^{5} GM/c^{2}$. The black lines correspond to the absence of DM, and the red, blue and green lines correspond to $\alpha=1.5$, $2.0$ and $7/3$, respectively. }
\label{pt_figures}
\end{figure*}

\begin{figure*}[htbp]
\centering
\includegraphics[height=5cm,width=7cm]{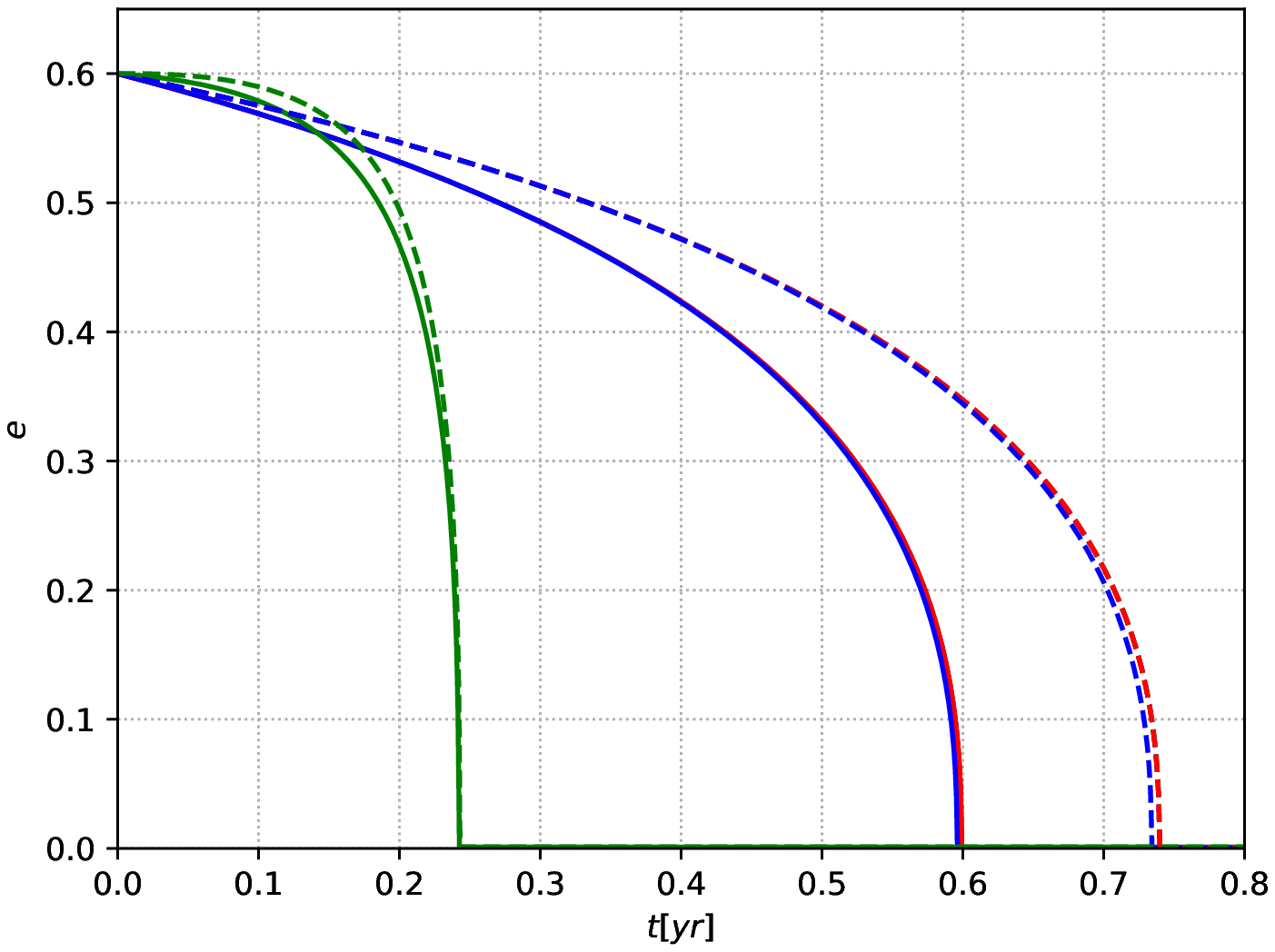}
\includegraphics[height=5cm,width=7cm]{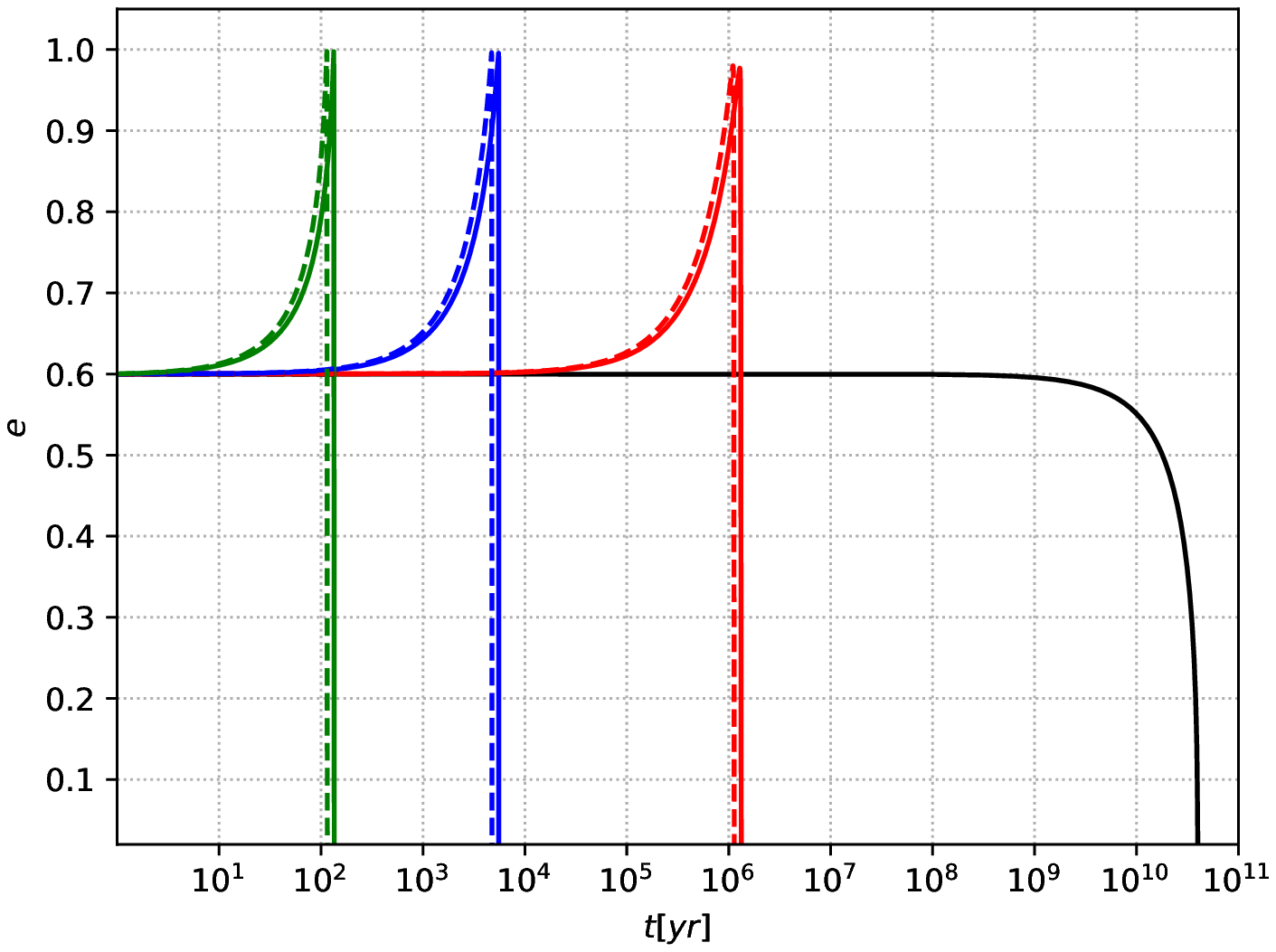}
\includegraphics[height=5cm,width=7cm]{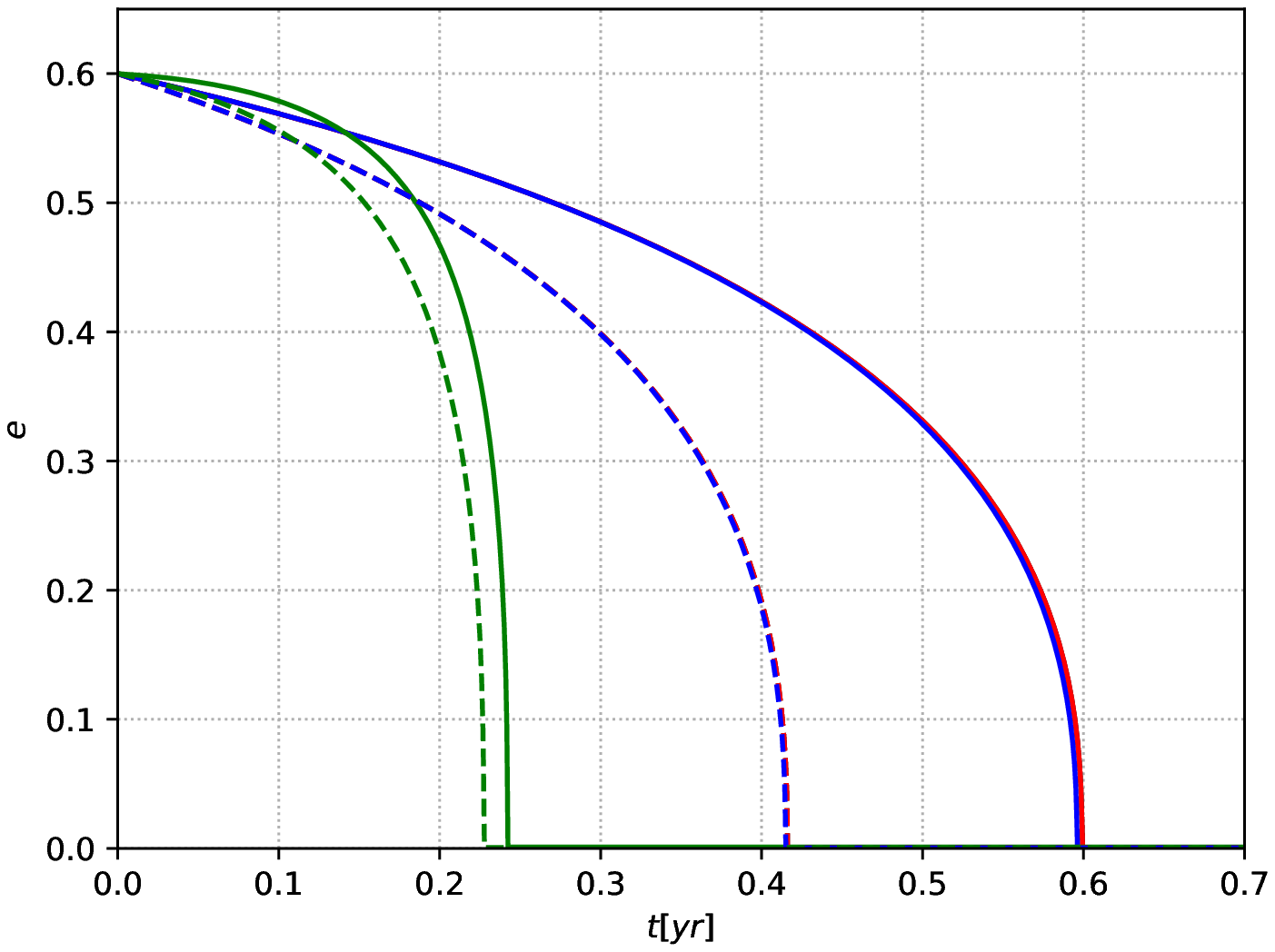}
\includegraphics[height=5cm,width=7cm]{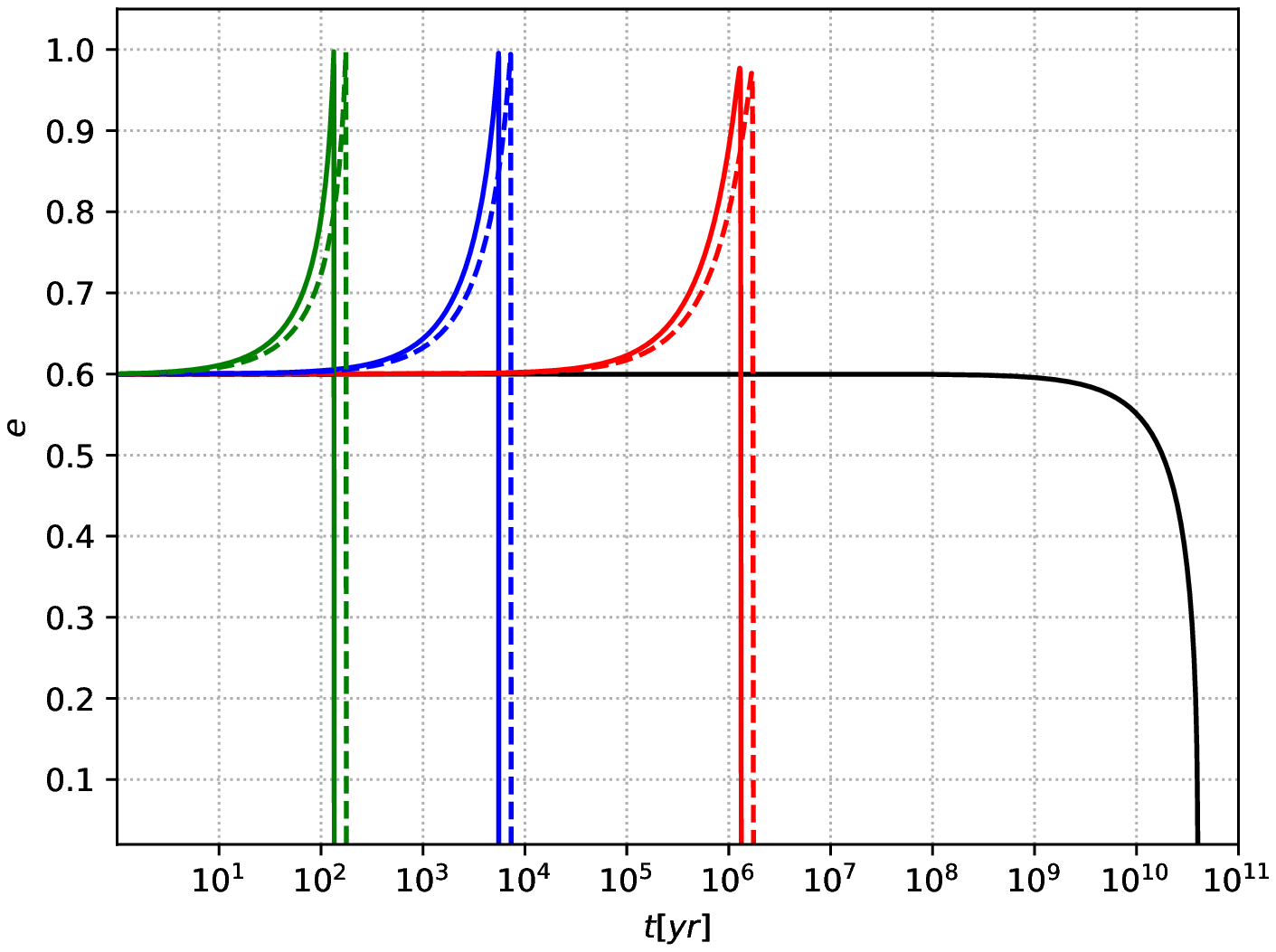}
\caption{The eccentricity $e$ of an IMRI evolves with time $t$ under different IMBH's masses. The horizontal axis is time $t$ with unit of year (yr), the vertical axis is the eccentricity $e$. We take the small compact object's mass as $10 M_{\odot}$ and the initial eccentricity $e=0.6$. The solid lines represent that the IMBH's mass is $10^{3} M_{\odot}$. The dashed lines represent that the IMBH's mass is $9\times10^{2} M_{\odot}$ for the upper panels and $1.2\times10^{3} M_{\odot}$ for the lower panels. The left panels: the initial $p$ is set to $200 GM/c^{2}$. The right panels: the relatively large initial is $p\simeq10^{5} GM/c^{2}$. The black lines correspond to the absence of DM, and the red, blue and green lines correspond to $\alpha=1.5$, $2.0$ and $7/3$, respectively. }
\label{et_figures}
\end{figure*}

Figure \ref{pt_figures} and \ref{et_figures} depicts the evolution of semi-latus rectum $p$ and eccentricity $e$ under different initial $p$ and different IMBH's masses. When the initial $p$ is relatively small, as shown in the left panels, only the denser DM spike with $\alpha=7/3$ has a significant effect on the evolution, and it is the same under different IMBH's masses. When the mass of IMBH increases slightly, the evolution will accelerate. When the initial $p$ is relatively lager, as shown in the right panels, the moderate DM spike can also accelerate the evolution, even under different IMBH's masses. When the mass of the IMBH increases slightly, the evolution will be delayed.

\subsection{DM halo without spike}

(1) NFW density profile

According to the derivation process of the subsection \ref{3-1}, for the NFW density profile, we can get the dynamical equations
\begin{equation}
\dot{p}=-\dfrac{64}{5}\dfrac{G^{3}\mu M^{2}}{c^{5}p^{3}}(1-e^{2})^{\dfrac{3}{2}}(1+\dfrac{7}{8}e^{2})-\dfrac{4G^{\dfrac{1}{2}}\mu\rho_{\rm{0}}R_{\rm{s}}^{3}p^{\dfrac{3}{2}} ln\Lambda}{M^{\dfrac{3}{2}}}(1-e^{2})^{\dfrac{3}{2}}\int_{0}^{2\pi}\dfrac{ (1+e\cos\phi)}{(1+2e\cos\phi+e^{2})^{\dfrac{3}{2}}[R_{\rm{s}}(1+e\cos\phi)+p]^{2}}d\phi ,
\label{p_t_NFW}
\end{equation}

\begin{equation}
\dot{e}=-\dfrac{304}{15}\dfrac{G^{3}\mu M^{2}}{c^{5}p^{4}}(1-e^{2})^{\dfrac{3}{2}}e(1+\dfrac{121}{304}e^{2})-\dfrac{4G^{\dfrac{1}{2}}\mu\rho_{\rm{0}}R_{\rm{s}}^{3}p^{\dfrac{1}{2}}ln\Lambda}{M^{\dfrac{3}{2}}}(1-e^{2})^{\dfrac{3}{2}}\int_{0}^{2\pi}\dfrac{(e+\cos\phi)(1+e\cos\phi)}{(1+2e\cos\phi+e^{2})^{\dfrac{3}{2}}[R_{\rm{s}}(1+e\cos\phi)+p]^{2}}d\phi .
\label{e_t_NFW}
\end{equation}

The semi-major axis $a$ and the semi-latus rectum $p$ satisfy the equation $ a=p/(1-e^{2}) $. According to the equations (\ref{p_t_NFW}) and (\ref{e_t_NFW}), using $a$ instead of $p$ to describe the dynamical equations, we can get 
\begin{equation}
\dot{e}=-\dfrac{304}{15}\dfrac{G^{3}\mu M^{2}}{c^{5}a^{4}}(1-e^{2})^{-\dfrac{5}{2}}e(1+\dfrac{121}{304}e^{2})-\dfrac{4G^{\dfrac{1}{2}}\mu\rho_{\rm{0}}R_{\rm{s}}^{3}a^{\dfrac{1}{2}}ln\Lambda}{M^{\dfrac{3}{2}}}(1-e^{2})^{2}\int_{0}^{2\pi}\dfrac{(e+\cos\phi)(1+e\cos\phi)}{(1+2e\cos\phi+e^{2})^{\dfrac{3}{2}}[R_{\rm{s}}(1+e\cos\phi)+a(1-e^{2})]^{2}}d\phi ,
\label{e_t_a_NFW}
\end{equation}

\begin{equation}
\dot{a}=-\dfrac{64}{5}\dfrac{G^{3}\mu M^{2}}{c^{5}a^{3}}(1-e^{2})^{-\dfrac{7}{2}}(1+\dfrac{73}{24}e^{2}+\dfrac{37}{96}e^{4})-\dfrac{4G^{\dfrac{1}{2}}\mu\rho_{\rm{0}}R_{\rm{s}}^{3}a^{\dfrac{3}{2}} ln\Lambda}{M^{\dfrac{3}{2}}}(1-e^{2})\int_{0}^{2\pi}\dfrac{ (1+e\cos\phi)}{(1+2e\cos\phi+e^{2})^{\dfrac{1}{2}}[R_{\rm{s}}(1+e\cos\phi)+a(1-e^{2})]^{2}}d\phi .
\label{a_t_NFW}
\end{equation}

(2) TF density profile

According to the derivation process of the subsection \ref{3-1}, for the TF density profile,  we can get the dynamical equations
\begin{equation}
\dot{p}=-\dfrac{64}{5}\dfrac{G^{3}\mu M^{2}}{c^{5}p^{3}}(1-e^{2})^{\dfrac{3}{2}}(1+\dfrac{7}{8}e^{2})-\dfrac{4G^{\dfrac{1}{2}}\mu\rho_{\rm{0}}p^{\dfrac{3}{2}} ln\Lambda}{kM^{\dfrac{3}{2}}}(1-e^{2})^{\dfrac{3}{2}}\int_{0}^{2\pi}\dfrac{\sin(\dfrac{kp}{1+e\cos\phi})}{(1+2e\cos\phi+e^{2})^{\dfrac{3}{2}}(1+e\cos\phi)}d\phi ,
\label{p_t_TF}
\end{equation}

\begin{equation}
\dot{e}=-\dfrac{304}{15}\dfrac{G^{3}\mu M^{2}}{c^{5}p^{4}}(1-e^{2})^{\dfrac{3}{2}}e(1+\dfrac{121}{304}e^{2})-\dfrac{4G^{\dfrac{1}{2}}\mu\rho_{\rm{0}}p^{\dfrac{1}{2}}ln\Lambda}{kM^{\dfrac{3}{2}}}(1-e^{2})^{\dfrac{3}{2}}\int_{0}^{2\pi}\dfrac{(e+\cos\phi)\sin(\dfrac{kp}{1+e\cos\phi})}{(1+2e\cos\phi+e^{2})^{\dfrac{3}{2}}(1+e\cos\phi)}d\phi .
\label{e_t_TF}
\end{equation}

The semi-major axis $a$ and the semi-latus rectum $p$ satisfy the equation $ a=p/(1-e^{2}) $. According to the equations (\ref{p_t_TF}) and (\ref{e_t_TF}), using $a$ instead of $p$ to describe the dynamical equations, we can get 
\begin{equation}
\dot{e}=-\dfrac{304}{15}\dfrac{G^{3}\mu M^{2}}{c^{5}a^{4}}(1-e^{2})^{-\dfrac{5}{2}}e(1+\dfrac{121}{304}e^{2})-\dfrac{4G^{\dfrac{1}{2}}\mu\rho_{\rm{0}}a^{\dfrac{1}{2}}ln\Lambda}{kM^{\dfrac{3}{2}}}(1-e^{2})^{2}\int_{0}^{2\pi}\dfrac{(e+\cos\phi)\sin[\dfrac{ka(1-e^{2})}{1+e\cos\phi}]}{(1+2e\cos\phi+e^{2})^{\dfrac{3}{2}}(1+e\cos\phi)}d\phi ,
\label{e_t_a_TF}
\end{equation}

\begin{equation}
\dot{a}=-\dfrac{64}{5}\dfrac{G^{3}\mu M^{2}}{c^{5}a^{3}}(1-e^{2})^{-\dfrac{7}{2}}(1+\dfrac{73}{24}e^{2}+\dfrac{37}{96}e^{4})-\dfrac{4G^{\dfrac{1}{2}}\mu\rho_{\rm{0}}a^{\dfrac{3}{2}} ln\Lambda}{kM^{\dfrac{3}{2}}}(1-e^{2})\int_{0}^{2\pi}\dfrac{\sin[\dfrac{ka(1-e^{2})}{1+e\cos\phi}]}{(1+2e\cos\phi+e^{2})^{\dfrac{1}{2}}(1+e\cos\phi)}d\phi .
\label{a_t_TF}
\end{equation}

(3) PI density profile

According to the derivation process of the subsection \ref{3-1}, for the PI density profile, we can get the dynamical equations
\begin{equation}
\dot{p}=-\dfrac{64}{5}\dfrac{G^{3}\mu M^{2}}{c^{5}p^{3}}(1-e^{2})^{\dfrac{3}{2}}(1+\dfrac{7}{8}e^{2})-\dfrac{4G^{\dfrac{1}{2}}\mu\rho_{\rm{0}}R_{\rm{c}}^{2}p^{\dfrac{5}{2}} ln\Lambda}{M^{\dfrac{3}{2}}}(1-e^{2})^{\dfrac{3}{2}}\int_{0}^{2\pi}\dfrac{1}{(1+2e\cos\phi+e^{2})^{\dfrac{3}{2}}[R_{\rm{c}}^{2}(1+e\cos\phi)^{2}+p^{2}]}d\phi ,
\label{p_t_PI}
\end{equation}

\begin{equation}
\dot{e}=-\dfrac{304}{15}\dfrac{G^{3}\mu M^{2}}{c^{5}p^{4}}(1-e^{2})^{\dfrac{3}{2}}e(1+\dfrac{121}{304}e^{2})-\dfrac{4G^{\dfrac{1}{2}}\mu\rho_{\rm{0}}R_{\rm{c}}^{2}p^{\dfrac{3}{2}}ln\Lambda}{M^{\dfrac{3}{2}}}(1-e^{2})^{\dfrac{3}{2}}\int_{0}^{2\pi}\dfrac{(e+\cos\phi)}{(1+2e\cos\phi+e^{2})^{\dfrac{3}{2}}[R_{\rm{c}}^{2}(1+e\cos\phi)^{2}+p^{2}]}d\phi .
\label{e_t_PI}
\end{equation}

The semi-major axis $a$ and the semi-latus rectum $p$ satisfy the equation $ a=p/(1-e^{2}) $. According to the equations (\ref{p_t_PI}) and (\ref{e_t_PI}), using $a$ instead of $p$ to describe the dynamical equations, we can get 
\begin{equation}
\dot{e}=-\dfrac{304}{15}\dfrac{G^{3}\mu M^{2}}{c^{5}a^{4}}(1-e^{2})^{-\dfrac{5}{2}}e(1+\dfrac{121}{304}e^{2}) -\dfrac{4G^{\dfrac{1}{2}}\mu\rho_{\rm{0}}R_{\rm{c}}^{2}a^{\dfrac{3}{2}}ln\Lambda}{M^{\dfrac{3}{2}}}(1-e^{2})^{3}\int_{0}^{2\pi}\dfrac{(e+\cos\phi)}{(1+2e\cos\phi+e^{2})^{\dfrac{3}{2}}[R_{\rm{c}}^{2}(1+e\cos\phi)^{2}+a^{2}(1-e^{2})^{2}]}d\phi ,
\label{e_t_a_PI}
\end{equation}

\begin{equation}
\dot{a}=-\dfrac{64}{5}\dfrac{G^{3}\mu M^{2}}{c^{5}a^{3}}(1-e^{2})^{-\dfrac{7}{2}}(1+\dfrac{73}{24}e^{2}+\dfrac{37}{96}e^{4}) $$$$
-\dfrac{4G^{\dfrac{1}{2}}\mu\rho_{\rm{0}}R_{\rm{c}}^{2}a^{\dfrac{5}{2}} ln\Lambda}{M^{\dfrac{3}{2}}}(1-e^{2})^{2}\int_{0}^{2\pi}\dfrac{1}{(1+2e\cos\phi+e^{2})^{\dfrac{1}{2}}[R_{\rm{c}}^{2}(1+e\cos\phi)^{2}+a^{2}(1-e^{2})^{2}]}d\phi .
\label{a_t_PI}
\end{equation}

\begin{figure*}[htbp]
\centering
\includegraphics[height=5cm,width=7cm]{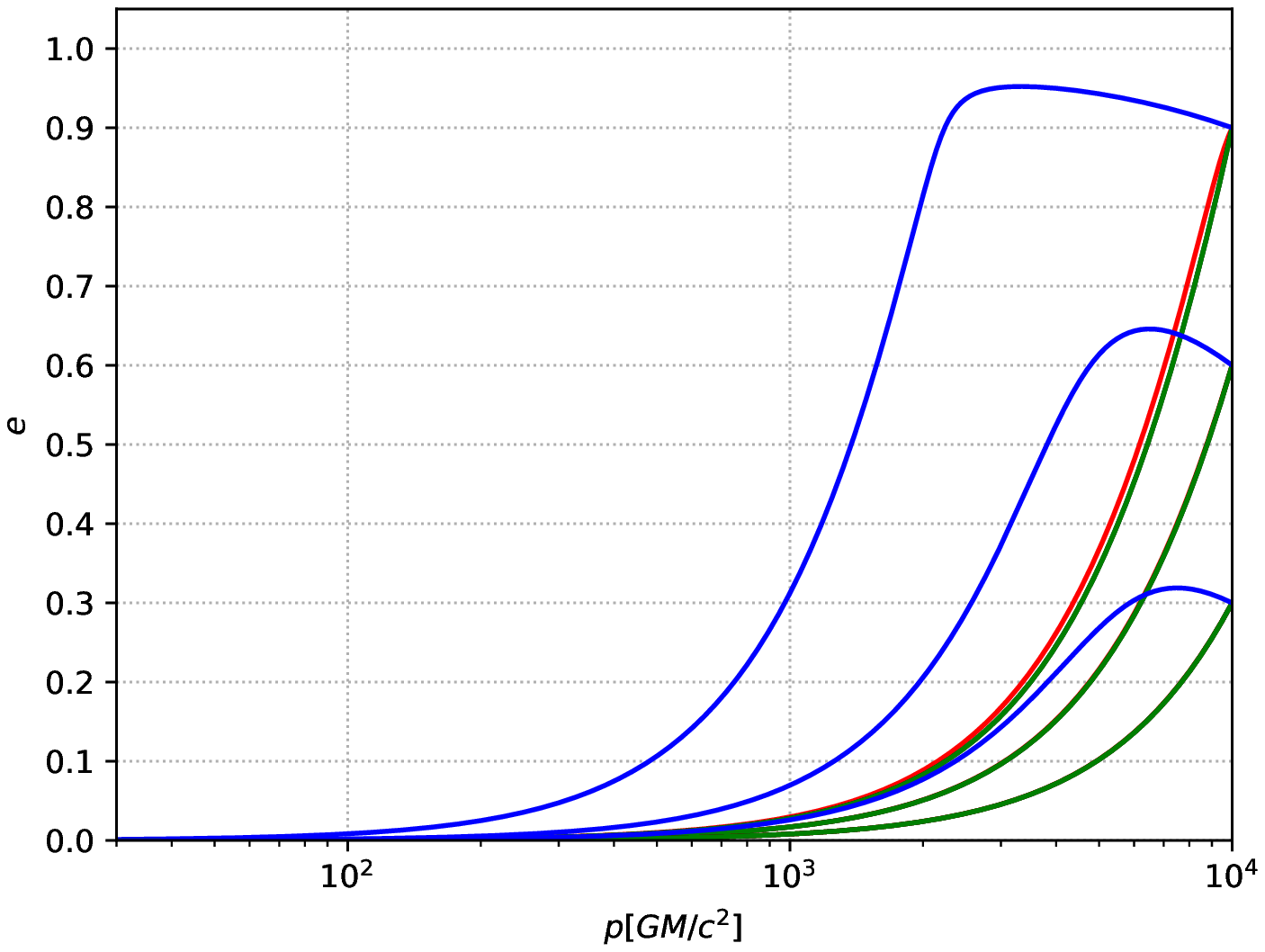}
\includegraphics[height=5cm,width=7cm]{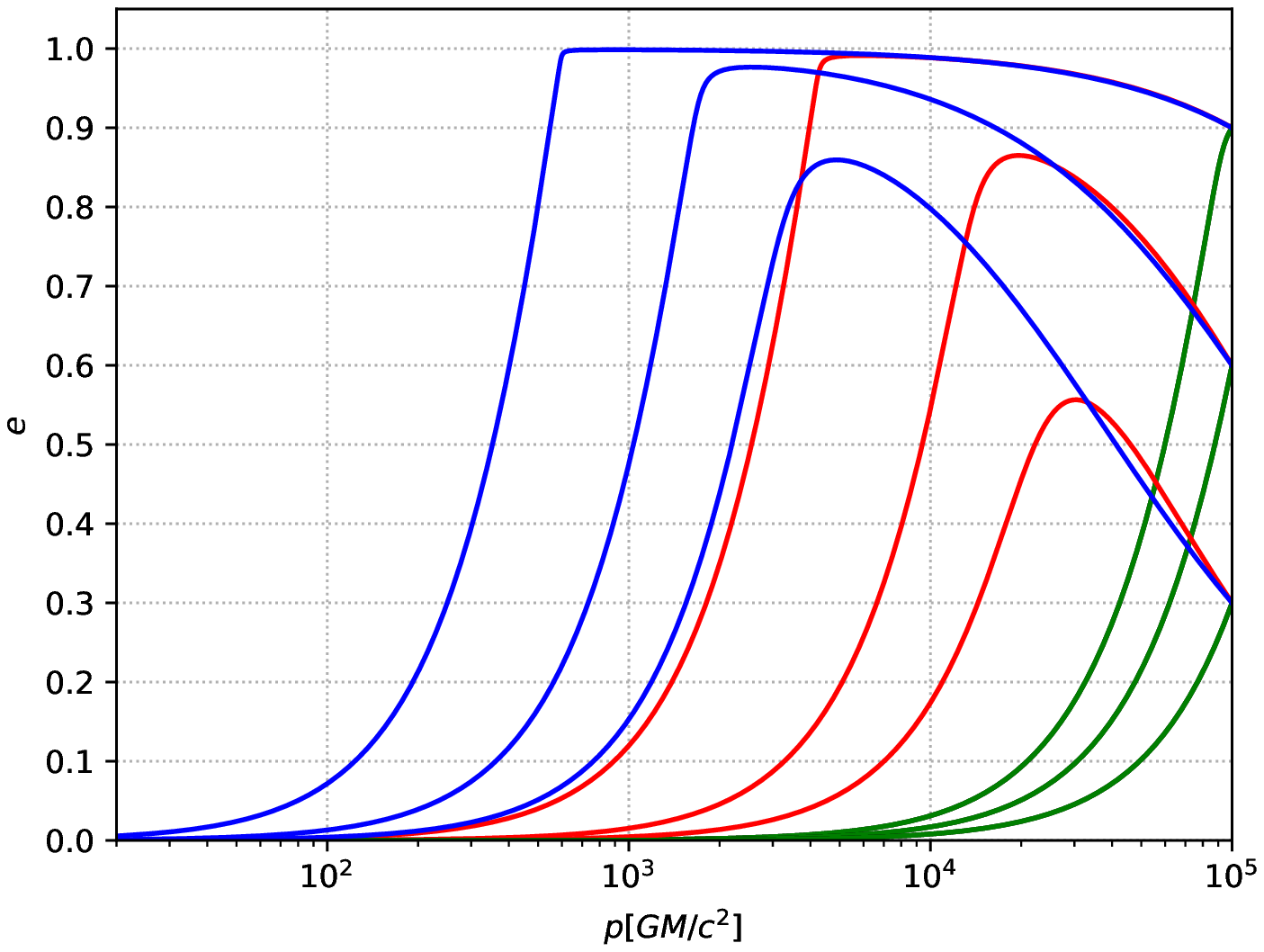}
\includegraphics[height=5cm,width=7cm]{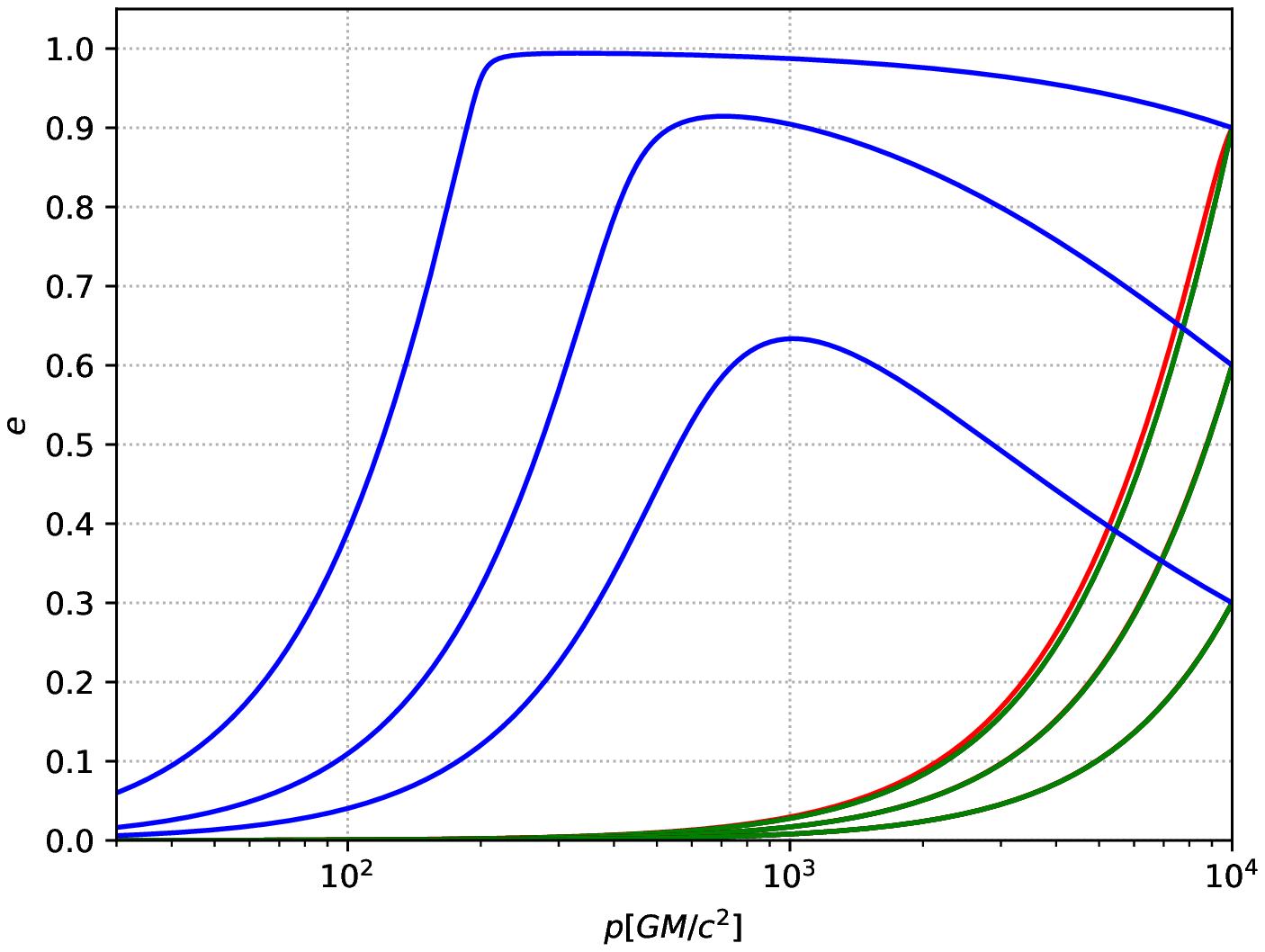}
\includegraphics[height=5cm,width=7cm]{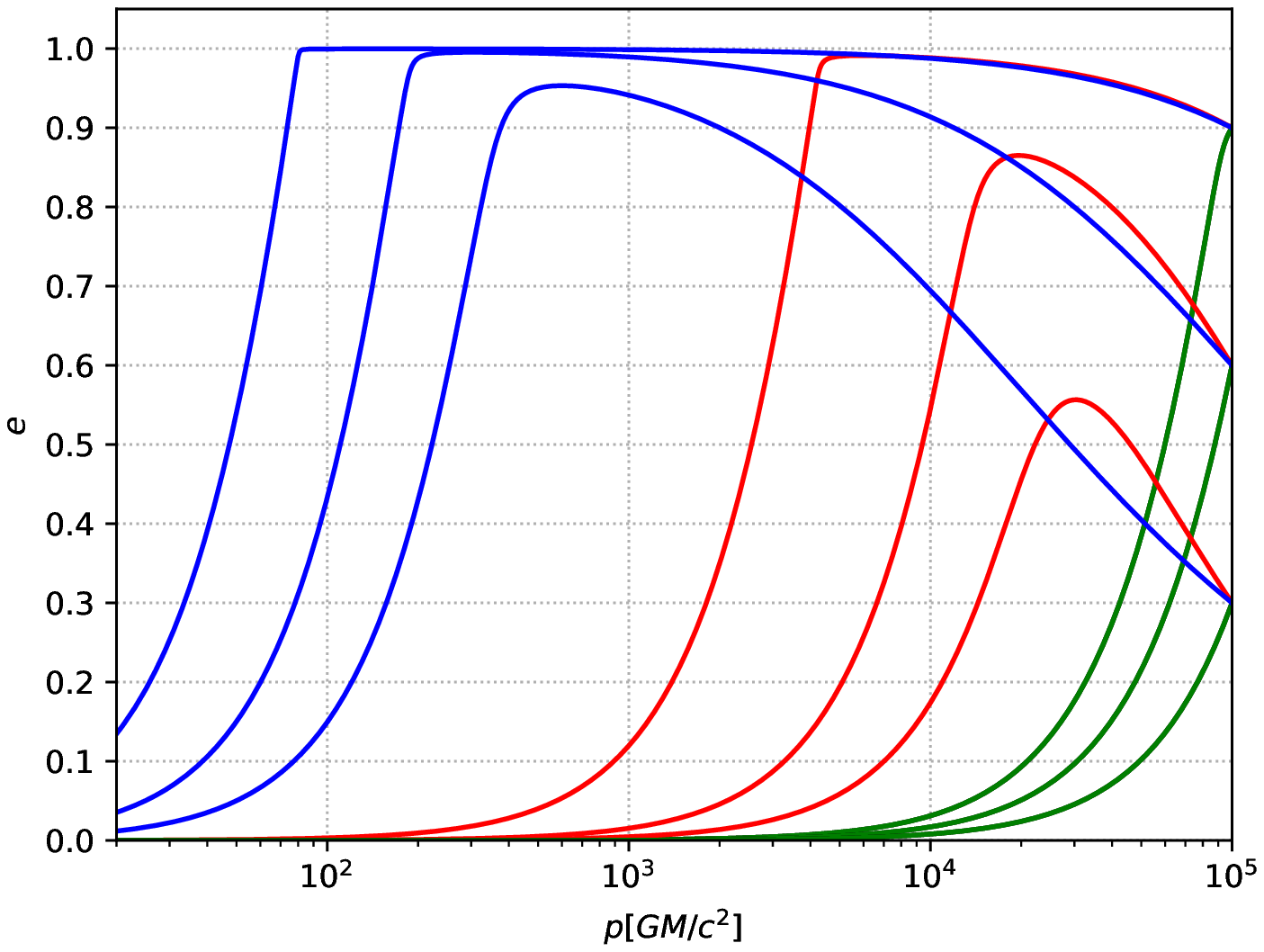}
\includegraphics[height=5cm,width=7cm]{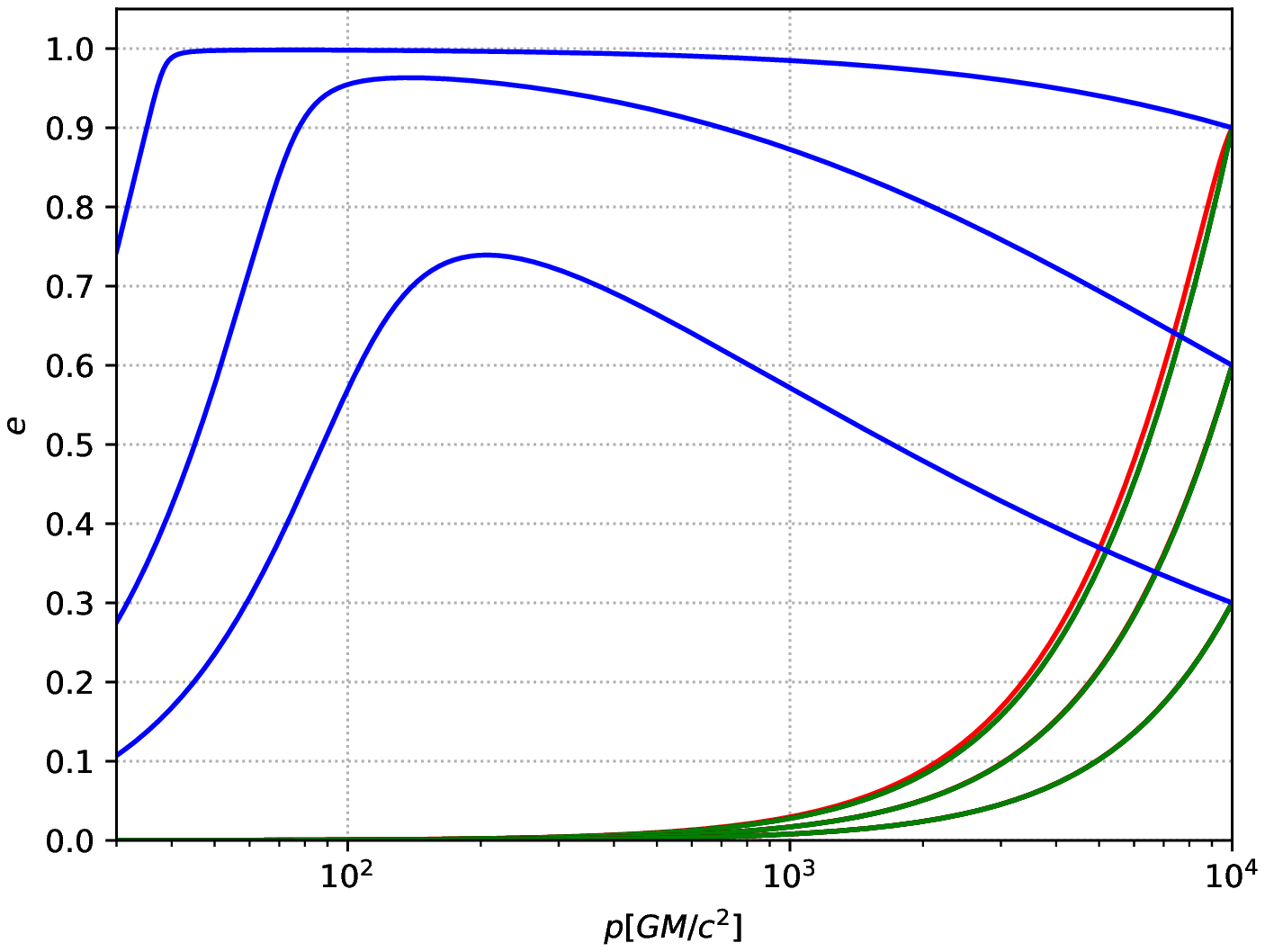}
\includegraphics[height=5cm,width=7cm]{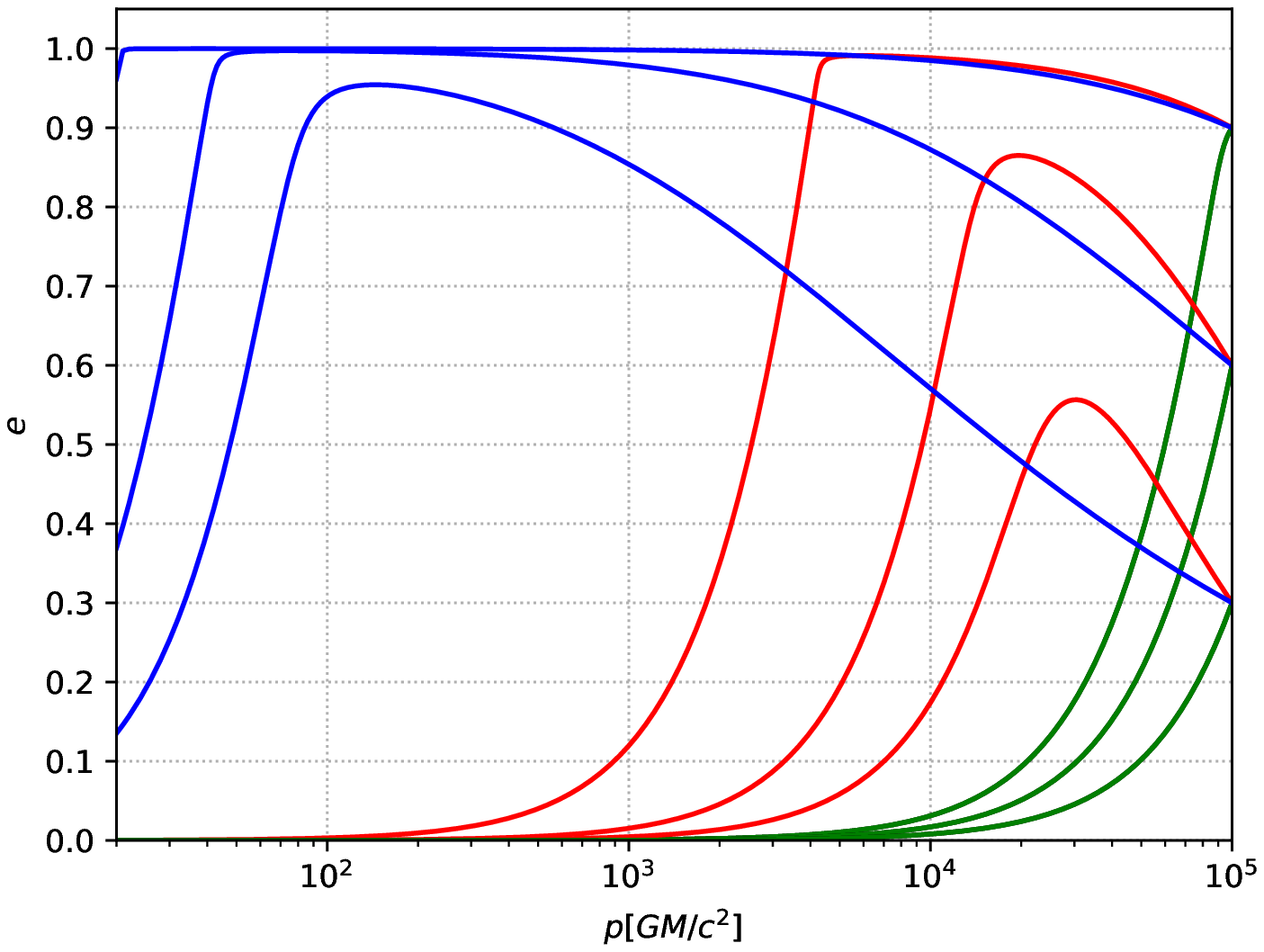}
\caption{The eccentricity $e$ of an IMRI evolves with the semi-latus rectum $p$ under different density profiles of DM. The horizontal axis is the semi-latus rectum $p$ with unit of $GM/c^{2}$, the vertical axis is the eccentricity $e$. In this figure, we take the small compact object's mass as $10 M_{\odot}$ and the IMBH's mass as $10^{3} M_{\odot}$. The blue lines correspond to the existence of DM spikes. The black lines correspond to the case without DM. The rest lines correspond to the cases without DM spikes (i.e., The red, green and yellow lines correspond to the NFW density profile, PI density profile and TF density profile, respectively. Here, the green, yellow and black lines overlap together). The left panels: the initial $p$ is set to $10^{4} GM/c^{2}$. The right panels: the relatively large initial is $p\simeq10^{5} GM/c^{2}$. These panels correspond to $\alpha=1.5$, $2.0$ and $7/3$ from top to bottom.}
\label{pe_4profiles_figures}
\end{figure*}

Figure \ref{pe_4profiles_figures} depicts the relation between $p$ and $e$ under different initial conditions and different DM profiles. When the initial $p$ is relatively small ($p=10^{4} GM/c^{2}$), as shown in the left panels , the three curves without DM spike (including NFW density profile, PI density profile and TF density profile) and the curves without DM are basically indistinguishable. However, the DM spikes with $\alpha=1.5$, $2.0$ and $7/3$ all can increase the eccentricity, and the larger the $\alpha$, the faster the eccentricity increases. For $\alpha=2.0$ and $7/3$, the eccentricity can even be close to $1$. When the initial $p$ is relatively large ($p\simeq10^{5} GM/c^{2}$), as shown in the right panels, in the case without DM spike, the curves of PI and TF density profiles overlap together completely, and the curves without DM are also completely consistent with them. And the NFW density profile can increase the eccentricity significantly. In the case of DM spike, for $\alpha=1.5$, $2.0$ and $7/3$, the eccentricity increases obviously, even approaching to $1$. When the eccentricity is near $1$, a portion of the curves for $\alpha=2.0$ and $7/3$ overlap together.

\begin{figure*}[htbp]
\centering
\includegraphics[height=5cm,width=7cm]{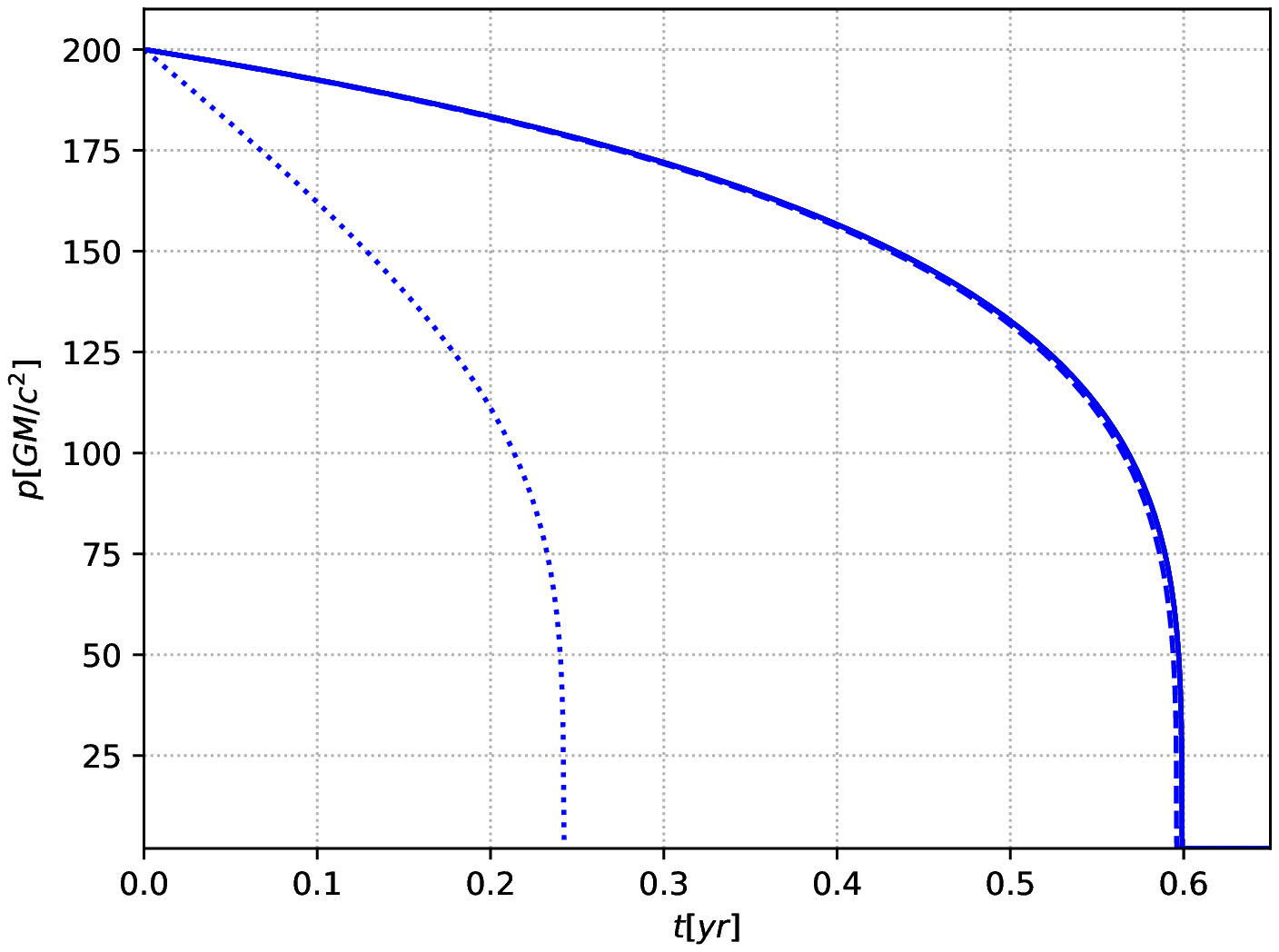}
\includegraphics[height=5cm,width=7cm]{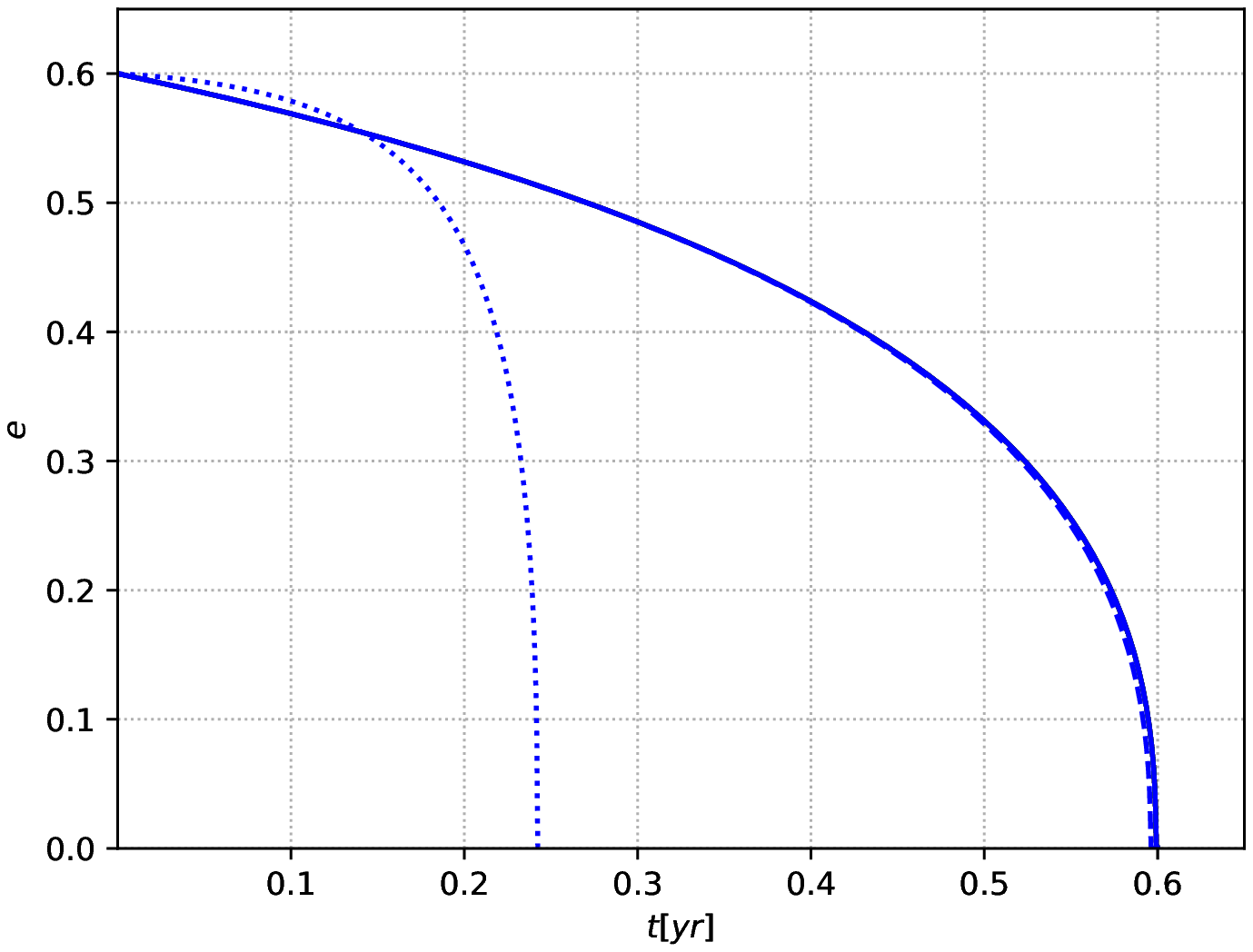}
\includegraphics[height=5cm,width=7cm]{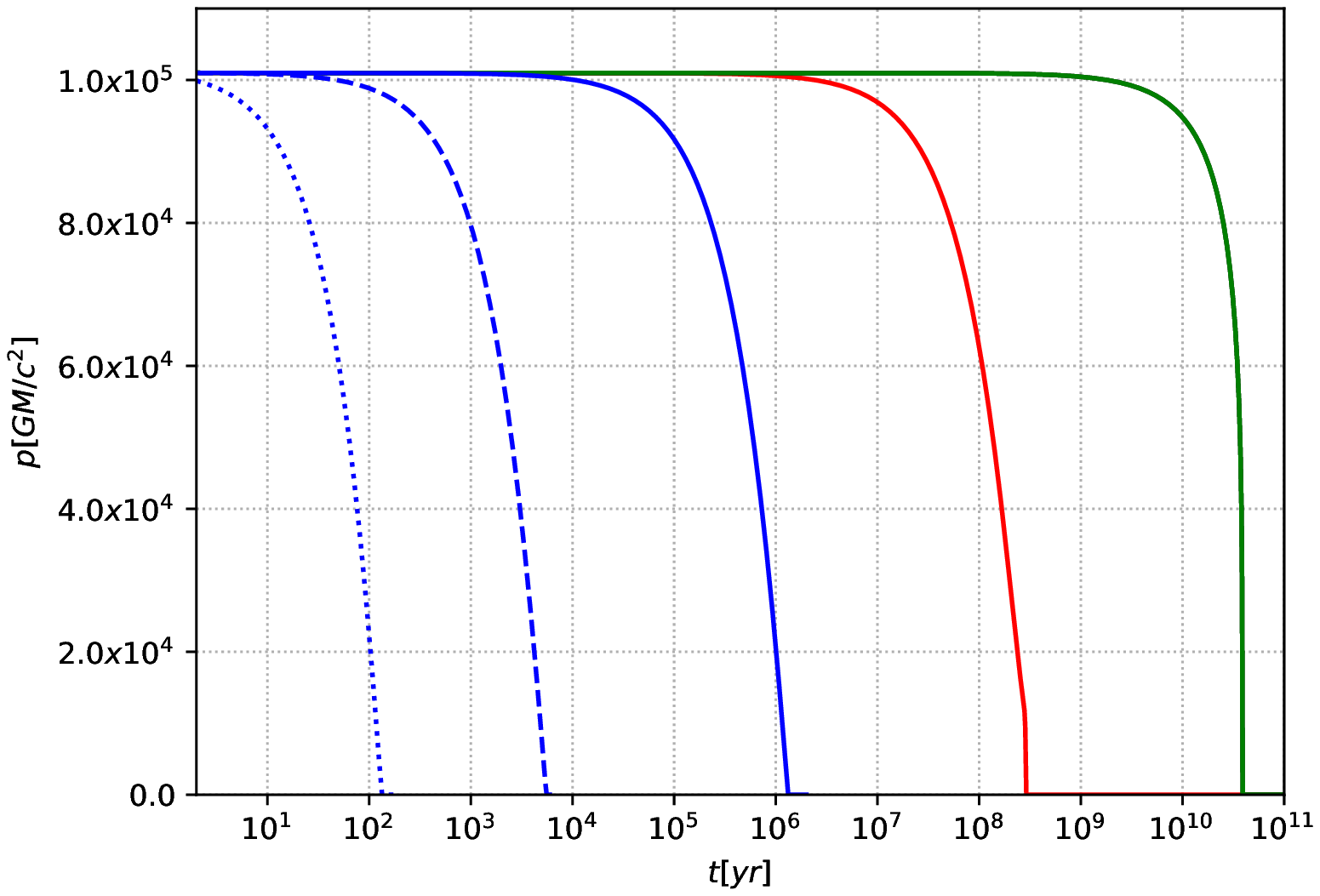}
\includegraphics[height=5cm,width=7cm]{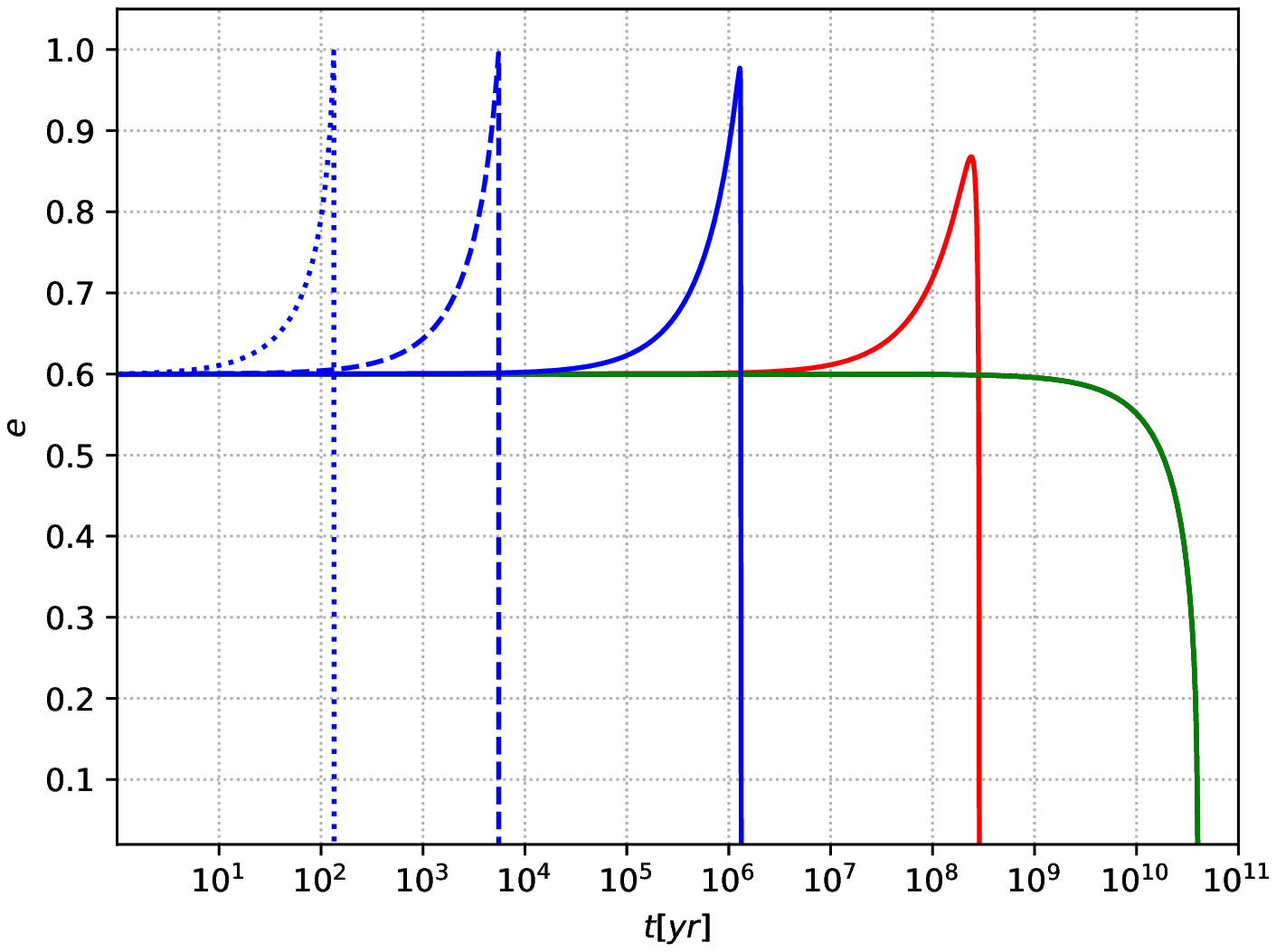}
\caption{The semi-latus rectum $p$ and the eccentricity $e$ of an IMRI evolve with  time $t$ under different initial $p$. The horizontal axis is time $t$ with unit of year (yr), the vertical axis is the semi-latus rectum $p$ with unit of $GM/c^{2}$ for the upper panels and the eccentricity $e$ for the lower panels. In this figure, we take the small compact object's mass as $10 M_{\odot}$, the IMBH's mass as $10^{3} M_{\odot}$ and the initial eccentricity $e=0.6$. The blue lines correspond to the existence of DM spike. The black lines correspond to the case without DM. The rest lines correspond to the cases without DM spikes (i.e., The red, green and yellow lines correspond to the NFW density profile, PI density profile and TF density profile, respectively. Here, the green, yellow and black lines overlap together). The solid, dashed and dotted blue lines correspond to $\alpha=1.5$, $2.0$ and $7/3$, respectively.The upper panels: the initial $p$ is set to $200 GM/c^{2}$. The lower panels: the relatively large initial is $p\simeq10^{5} GM/c^{2}$.}
\label{et_4profiles_figures}
\end{figure*}

Figure \ref{et_4profiles_figures} describes the evolution of $p$ and $e$ under different initial conditions and different DM profiles. When the initial $p$ is relatively small ($p=200 GM/c^{2}$), as shown in the upper panels, the DM profiles without spike (including NFW density profile, PI density profile and TF density profile) and the case without DM are basically indistinguishable from profile of DM spike with $\alpha=1.5$ and $2.0$ in terms of their impact on evolution. Only the denser DM spike with $\alpha=7/3$ influences the evolution significantly. When the initial $p$ is relatively large ($p\simeq10^{5} GM/c^{2}$), as shown in the lower panels, no matter whether there is DM spike, it will accelerate the evolution. But the presence of DM spike has a greater impact on the evolution, and when $\alpha$ is larger, the evolution will be faster.

\section{SUMMARY AND CONCLUSIONS}
\label{conclusion}

In this work, we have considered the effect of DM and IMBH's mass on the eccentricity for an IMRI system. Specifically, we have considered the influence of the DM spike on the eccentricity under the same mass of the center IMBH, the change of different IMBH masses to the eccentricity when the DM spike exists, and the change of DM halo to the eccentricity under the absence of DM spike. We found that: $(1)$ the mass of the center IMBH can be measured by observing the change of orbital eccentricity of the stellar massive BH at different scales, and the measurable mass will meet a certain range; $(2)$ by measuring the orbital eccentricity of the stellar massive BH for an IMRI, it is possible to study the DM model at the scale of $10^{5} GM/c^{2}$.

When the DM spike are present, the eccentricity is increased. For the denser DM spike with $\alpha=7/3$ and a larger initial $p$, the eccentricity will increase obviously, which is consistent with the results of \citep{2019PhRvD.100d3013Y}. This indicates that the result is no different from the power-law distribution by using the more accurate profile of DM spike. In the presence of DM spike, we found that by adjusting the IMBH's mass, the orbital eccentricity of the stellar massive BH would change accordingly. Specifically, when the mass of the center IMBH increases, the enhancement effect of the eccentricity will decrease significantly; when the IMBH's mass decreases, the enhancement effect of the eccentricity will increase obviously. This suggests that by observing the eccentricity of the stellar massive BH, it is possible to infer the mass of the center BH. Of course, measuring the eccentricity at different scales, the center BH's mass that can be detected is different.

Next, we have obtained the magnitude of the eccentricity's enhancement of the stellar BH in the presence and absence of DM spike. We found that when there is a DM spike near the IMBH, the eccentricity has a significant enhancement effect at the scale of $20 GM/c^{2}\sim 10^{5} GM/c^{2}$; but when there is no DM spike and only the DM halo is considered, the eccentricity increases obviously at the scale of about $10^{5} GM/c^{2}$, and the increase is much smaller than in the case with DM spike. This shows that by measuring the eccentricity of the stellar BH at the scale of $10^{5} GM/c^{2}$, it is possible to investigate the distribution of DM in the vicinity of the center BH, so as to study the DM model more deeply.

In future work, we will calculate the enhancement effect of different BH models and their properties on the eccentricity for an IMRI, so as to study the BH model and essence.

\acknowledgments
We thank the anonymous reviewer for a constructive report that has significantly improved this paper. We acknowledge the financial support from the National Natural Science Foundation of China under grants No. 11503078, 11573060 and 11661161010.

\end{document}